\pdfoutput=1
\documentclass[acmlarge]{acmart}

\renewcommand\footnotetextcopyrightpermission[1]{}
\usepackage{booktabs}

\usepackage[ruled]{algorithm2e}

\usepackage[utf8]{inputenc}

\usepackage{mathtools,amsmath,graphicx,bm,tikz,url,rotating}
\usetikzlibrary{arrows,positioning,shapes}
\usepackage{multirow, siunitx}

\usepackage{soul}

\usepackage{xcolor, pgfplots}

\newcommand{\tab}{\hspace{2mm}} % used in Item Table

\newcommand{\ci}[3]{ $\underset{(#1\;\;#3)}{#2}$}

\newcommand{\msdd}[2]{ $#1\; (#2$)}
\newcommand{\msddb}[2]{ #1 & (#2)}

\newcommand{\superscript}[1]{\ensuremath{^{\textrm{#1}}}}

\definecolor{cusa}{HTML}{9e0142}
\definecolor{cspain}{HTML}{D53E4F}
\definecolor{cgermany}{HTML}{F46D43}
\definecolor{cbrazil}{HTML}{FDAE61}
\definecolor{crussia}{HTML}{FEE08B}
\definecolor{cmexico}{HTML}{d1ed45}
\definecolor{cindia}{HTML}{abdda4} 
\definecolor{cindonesia}{HTML}{66c2a5}
\definecolor{cphilipines}{HTML}{3288bd}
\definecolor{cvenezuela}{HTML}{5E4FA2}

\SetAlFnt{\small}
\SetAlCapFnt{\small}
\SetAlCapNameFnt{\small}
\SetAlCapHSkip{0pt}
\IncMargin{-\parindent}

\setcopyright{acmlicensed}

\acmDOI{0000001.0000001}

\begin{document}

\setstcolor{red}

\title[The Multidimensional Crowdworker Motivation Scale]{Measuring Motivations of Crowdworkers:\\The Multidimensional Crowdworker Motivation Scale}

\author{Lisa Posch}
\affiliation{
  \institution{GESIS -- Leibniz Institute for the Social Sciences}
  \streetaddress{Unter Sachsenhausen 6-8}
  \city{Cologne}
  \postcode{50667}
  \country{Germany}
}
\affiliation{
  \institution{Graz University of Technology}
  \streetaddress{Inffeldgasse 16c}
  \city{Graz}
  \postcode{8010}
  \country{Austria}
}
\email{lisa.posch@gesis.org}

\author{Arnim Bleier}
\affiliation{
  \institution{GESIS -- Leibniz Institute for the Social Sciences}
  \streetaddress{Unter Sachsenhausen 6-8}
  \city{Cologne}
  \postcode{50667}
  \country{Germany}
}
\email{arnim.bleier@gesis.org}

\author{Clemens Lechner}
\affiliation{
  \institution{GESIS -- Leibniz Institute for the Social Sciences}
  \streetaddress{B2,1}
  \city{Mannheim}
  \postcode{68159}
  \country{Germany}
 }
\email{clemens.lechner@gesis.org}

\author{Daniel Danner}
\affiliation{
  \institution{University of Applied Labour Studies}
  \streetaddress{Seckenheimer Landstraße 16}
  \city{Mannheim}
  \postcode{68163}
  \country{Germany}
}
\email{daniel.danner@arbeitsagentur.de}

\author{Fabian Flöck}
\affiliation{
  \institution{GESIS -- Leibniz Institute for the Social Sciences}
  \streetaddress{Unter Sachsenhausen 6-8}
  \city{Cologne}
  \postcode{50667}
  \country{Germany}
}
\email{fabian.floeck@gesis.org}

\author{Markus Strohmaier}
\affiliation{
  \institution{RWTH Aachen University}
  \streetaddress{Theaterplatz 14}
  \city{Aachen}
  \postcode{52062}
  \country{Germany}
}
\affiliation{
  \institution{GESIS -- Leibniz Institute for the Social Sciences}
  \streetaddress{Unter Sachsenhausen 6-8}
  \city{Cologne}
  \postcode{50667}
  \country{Germany}
}
\email{markus.strohmaier@humtec.rwth-aachen.de}

\begin{abstract}
Crowd employment is a new form of short-term and flexible employment which has emerged during the past decade.
In order to understand this new form of employment, it is crucial to illuminate the underlying motivations of the workforce involved in it. This paper introduces the Multidimensional Crowdworker Motivation Scale (MCMS), a scale for measuring the motivation of crowdworkers on micro-task platforms. The MCMS is theoretically grounded in self-determination theory and tailored specifically to the context of paid crowdsourced micro-labor. 
The scale measures the motivation of crowdworkers along six motivational dimensions, ranging from amotivation to intrinsic motivation.
We validated the MCMS on data collected in ten countries and three income groups. Factor analyses demonstrated that the MCMS's six dimensions showed good model fit, validity, and reliability.
Furthermore, our measurement invariance tests showed that motivations measured with the MCMS are comparable across countries and income groups, and we present a first cross-country comparison of crowdworker motivations. This work constitutes an important first step towards understanding the motivations of the international crowd workforce.
\end{abstract}


\maketitle

\thispagestyle{empty}

\renewcommand{\shortauthors}{L. Posch et al.}

\section{Introduction}

During the past decade, crowd employment has emerged as a new form of short-term and flexible employment. As such, crowd employment is part of a wider trend in industrial societies toward increasingly flexible work arrangements that are characterized by short-term, market based contracts \cite{kalleberg2009precarious, hewison2013precarious}.
Crowd employment has been defined as a type of employment which ``uses an online platform to enable organisations or individuals to access an indefinite and unknown group of other organisations or individuals to solve specific problems or to provide specific services or products in exchange for payment'' \cite{mandl2015new}. While this definition is similar to the concept of crowdsourcing \cite{crowdsourcing_definition}, it explicitly includes only those activities that are performed in exchange for payment.

One type of crowd employment platforms are micro-task platforms such as Amazon Mechanical Turk\footnote{http://www.mturk.com/} (AMT) or CrowdFlower\footnote{http://www.crowdflower.com/}. 
On micro-task platforms, crowdworkers are paid on a per-task basis, and a single task usually pays only a few cents upon completion.  
The micro tasks offered to workers on these platforms are also called ``human intelligence tasks'' 
and typically require workers to solve problems that are easy to solve for humans but hard to solve for computers. This characteristic led Amazon Mechanical Turk to coin the term ``artificial artificial intelligence'' to describe this type of work. Typical micro tasks include classification and tagging of text or images, audio and image transcription, and validating addresses of companies on the web. Also more complex tasks such as editing text documents \cite{bernstein2015soylent}, ontology alignment \cite{sarasua2012crowdmap} and the evaluation of unsupervised machine learning algorithms (e.g. \cite{chang2009reading, foulds2013stochastic, posch2015polylingual}) have been successfully deployed on micro-task platforms. Anyone, regardless of geographical location or education, can perform micro tasks -- the only necessary requirement is having access to the Internet.

The emergence of crowd employment and a general trend towards more flexible and shorter-term employment have given rise to policy discussions on social protection and working conditions of crowdworkers (e.g. \cite{europarl,euroagenda,codagnone2016future,felstiner2011working}).
One ongoing discussion is whether crowd employment is to be considered ``work'' at all, or whether it is mostly considered a spare-time activity by many workers, meaning that remuneration plays only a minor role for them \cite{europarl}.
Estimates of the hourly wage achievable on popular micro-task platforms lie between under US\$1 and around US\$5 \cite{berg2016income, berg2018digital, ross2010crowdworkers, horton2010labor, khanna2010evaluating}. 
While this amount is above the minimum wage in some countries, in many high-income countries it is far below the wage of any traditional job. 
Despite this, the rise of crowd employment is an international phenomenon which does not exclude high-income countries.
Understanding the underlying motivations of the international, ``indefinite and unknown group'' of crowdworkers is therefore crucial for understanding this new form of employment. 

Although there has been some research on the motivation of crowdworkers, 
there is currently no theoretically founded scale for comprehensively measuring motivations of crowdworkers in different countries.
So far, the question of what motivates people across the world to participate in micro-task crowdwork remains largely open.

This paper lays the groundwork for understanding the motivations of the international crowd workforce by introducing the Multidimensional Crowdworker Motivation Scale (MCMS) and presenting a case study conducted on a large sample of crowdworkers from ten different countries. The MCMS is theoretically grounded in self-determination theory (SDT) and tailored specifically to the context of paid crowdsourced micro-labor. Most items in the MCMS are based on items from existing SDT-based motivation scales developed for the traditional work context, which we adapted to the idiosyncrasies of work on micro-task platforms.

The main contributions of this paper are (1) an evaluation of two existing SDT-based work motivation scales developed for the traditional work context with respect to their suitability for micro-task crowdwork; (2) the development of the Multidimensional Crowdworker Motivation Scale (MCMS) that draws from these existing scales but refines them and adapts them to the context of micro tasks; (3) a validation of the MCMS in ten countries and three income groups; (4) an evaluation of the comparability of motivations measured with the MCMS across countries and across income groups; and (5) a first analysis of differences in crowdworker motivations across these countries and income groups. 
To the best of our knowledge, the MCMS is the first motivation scale developed specifically for the context of crowdsourced micro-labor that offers a comprehensive representation of the motivational dimensions according to SDT. 
Furthermore, it is the first motivation scale for the crowdworking domain that is validated across multiple countries and income groups. 

The paper is structured in the following way: Section~\ref{sec:related_work} gives a short overview of the different types of motivation as conceptualized by self-determi\-nation theory and reviews existing SDT-based work motivation scales. Furthermore, it gives an overview of related work on the motivations of crowdworkers on micro-task platforms. 
In Section~\ref{sec:existing}, we evaluate to what extent SDT-based motivation scales developed for the traditional work context can be successfully applied for measuring crowdworker motivations, and we show the need for a work motivation scale adapted to the idiosyncrasies of the micro-task context. Section~\ref{sec:development} describes the process of developing the MCMS and Section~\ref{sec:validation} presents a validation of the MCMS in ten countries and three income groups. In Section~\ref{sec:invariance}, we demonstrate the cross-country and cross-income group comparability of motivations measured with the MCMS. Section~\ref{sec:results} presents a first cross-country and cross-income group comparison of crowdworker motivations. Finally, Section~\ref{sec:conclusion} concludes this work and discusses the scale's limitations as well as directions for future research.

\section{Related Work}
\label{sec:related_work}

\textbf{Self-Determination Theory and Work Motivation. }
Self-determi\-nation theory (SDT) is a theory of human motivation that was developed by Deci and Ryan \cite{deci1980empirical, deci868, deci2000and}. 
The theory specifies three general kinds of motivation that are hypothesized to lie along a continuum of self-determination: \emph{amotivation}, \emph{extrinsic motivation} and \emph{intrinsic motivation}. At the one extreme of the continuum lies \emph{amotivation}, which completely lacks self-determination; at the other extreme lies \emph{intrinsic motivation}, which is completely self-determined \cite{gagne2005self}. 
Between these extremes lies \emph{extrinsic motivation}, which is further split up into subtypes with varying degrees of internalisation: \emph{external regulation}, \emph{introjected regulation}, \emph{identified regulation} and \emph{integrated regulation}.

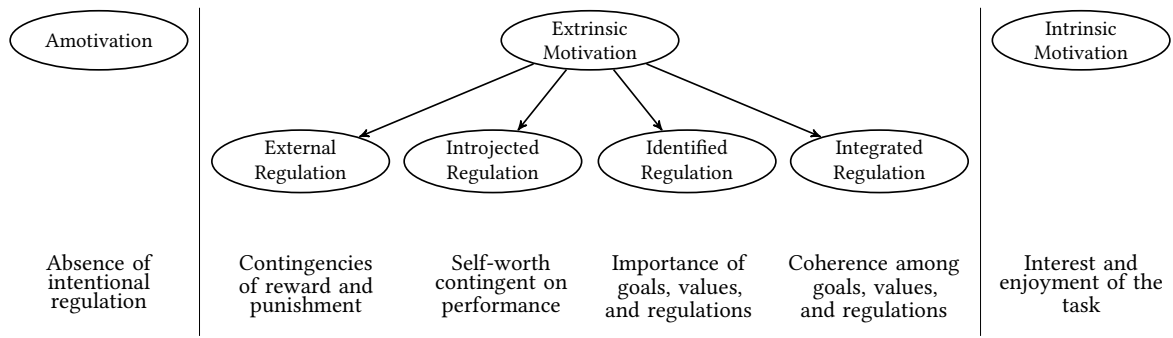
\begin{figure}
\centering
\resizebox{\columnwidth}{!}{
\begin{tikzpicture}[auto,node distance=.20cm,
    latent/.style={ellipse,draw, thick,inner sep=0pt,minimum height =10mm, minimum width = 30mm,align=center},
    paths/.style={->, thick, >=stealth'},
    mytxt/.style={anchor=north, text width=3.2cm, align=center, font=\fontsize{11pt}{9pt}\selectfont},
]

\node [latent] (EXTERNAL) at (0,0) {Extrinsic\\Motivation};

\node [latent] (EXTERNAL_REG) [below left = 1.3cm and 2.75cm of EXTERNAL] {External\\Regulation};
\node [latent] (INTROJECTED_REG) [below left = 1.3cm and -.5cm of EXTERNAL] {Introjected\\Regulation};
\node [latent] (INTEGRATED_REG) [below right = 1.3cm and 2.75cm of EXTERNAL] {Integrated\\Regulation};
\node [latent] (IDENTIFIED_REG) [below right = 1.3cm and -.5cm of EXTERNAL] {Identified\\Regulation};

\draw [paths] (EXTERNAL) to node { } (EXTERNAL_REG);
\draw [paths] (EXTERNAL) to node { } (INTROJECTED_REG);
\draw [paths] (EXTERNAL) to node { } (INTEGRATED_REG);
\draw [paths] (EXTERNAL) to node { } (IDENTIFIED_REG);

\node [latent] (AMOTIVATION) at (-8.3,0) {Amotivation};
\node [latent] (INTRINSIC) at (8.3,0) {Intrinsic\\Motivation};

\draw (-6.6,.55) -- (-6.6,-5);
\draw (6.6,.55) -- (6.6,-5);

\node[mytxt] at (-8.3,-3.5) {Absence of\\intentional regulation};
\node[mytxt] at (-4.8,-3.5) {Contingencies of reward and punishment};
\node[mytxt] at (-1.5,-3.5) {Self-worth contingent on performance};
\node[mytxt] at (1.5,-3.5) {Importance of goals, values, and regulations};
\node[mytxt] at (4.8,-3.5) {Coherence among goals, values, and regulations};
\node[mytxt] at (8.3,-3.5) {Interest and\\enjoyment of the\\task};

\end{tikzpicture}
}
\caption{\textbf{Types of motivation.} This figure shows the different types of motivation along the self-determination continuum hypothesized by SDT. The figure was adopted from Gagné and Deci \cite{gagne2005self}.} 
\label{fig:sdt}
\end{figure}

Figure \ref{fig:sdt} (adopted from \cite{gagne2005self}) shows the types of motivation as specified by SDT. 
\emph{Amotivation} is the absence of motivation, a state of acting passively or not intending to act all. 
\emph{External regulation} is the least self-determined form of extrinsic motivation. Individuals motivated by external regulation act in order to obtain rewards or avoid punishments. 
\emph{Introjected regulation} refers to a form of partially internalized extrinsic motivation which aims at the avoidance of guilt or at attaining feelings of worth \cite{deci2002overview}.
\emph{Identified regulation} is a form of extrinsic motivation with a high degree of perceived autonomy, where the action is in alignment with the individual's personal goals.
\emph{Integrated regulation} is the most self-determined form of extrinsic motivation and stems from evaluated identifications that are in alignment with self-endorsed values, goals and needs \cite{deci2002overview}.
The most self-determined form of motivation is \emph{intrinsic motivation}. This form of motivation is non-instrumental and people act freely, driven by interest and enjoyment inherent in the action \cite{ryan2000intrinsic}.

SDT hypothesizes that individuals may internalize an initially external regulation, which then becomes more self-determined. Regulations can be internalized in different ways, depending on the extent to which the individual has integrated it with his or her sense of self \cite{deci2002handbook}. For example, an activity could initially be externally regulated because it is not perceived as enjoyable by the individual. However, when this activity becomes valued by the person, for example because it is perceived to be important for his or her personal goals, the regulation for this behavior is internalized (in this case, becoming identified regulation).

While SDT postulates that the different types of internalization fall along a continuum structure, empirical evidence for this continuum hypothesis is inconsistent (e.g. \cite{chemolli2014evidence, howard2016using, litalien2015motivation, guay2015application}). For example, Chemolli and Gagné \cite{chemolli2014evidence} showed that motivations differ more in \emph{kind} than in \emph{degree}, and that SDT-based motivation scales are best represented by multidimensional models. Howard et al. \cite{howard2016using} found evidence of a global factor measuring the quantity of self-determination, but they also found that each of the motivation types provided unique information beyond the quantity of self-determination.

Several work motivation scales for the traditional employment context
have been developed based on SDT.
The first SDT-based work motivation scale was a French scale developed by Blais et al. \cite{blais1993work}. Tremblay et al. \cite{weims} translated this scale into English and conducted an evaluation in different work environments. The resulting Work Extrinsic and Intrinsic Motivation Scale (WEIMS) measures six factors: amotivation, the four external regulation subtypes and intrinsic motivation. 
Gagné et al. \cite{gagne2010motivation} created the Motivation at Work Scale (MAWS), a scale that measures the four factors external regulation, introjected regulation, identified regulation and intrinsic motivation. The MAWS was validated in French and in English and was partly based on the scale developed by Blais et al. \cite{blais1993work}.

Later, Gagné et al. \cite{mwms} developed the Multidimensional Work Motivation Scale (MWMS). The MWMS was validated in seven languages and nine countries and does not include any items from the MAWS.
The MWMS measures six first-order factors (amotivation, material external regulation, social external regulation, introjected regulation, identified regulation and intrinsic motivation) and one second-order factor (external regulation).
Work motivation scales such as MWMS, MAWS and WEIMS investigate motivations at the domain level of analysis, meaning that they measure the general motivation to perform a job as opposed to specific tasks within a job.

\textbf{Crowdworker Motivation on Micro-Task Platforms.} 
Compared to work motivation in the traditional employment context, research on motivation in the micro-task context is still scarce and scattered. Most research investigating the motivations of workers on micro-task platforms has focused on the platform Amazon Mechanical Turk (AMT). Consequently, most studies have focused on American and Indian crowdworkers, which constitute the vast majority of workers on AMT\footnote{American and Indian crowdworkers currently constitute over 80\% of the worker population on ATM (also see \url{http://demographics.mturk-tracker.com/##/countries/all}).} \cite{ipeirotis2010analyzing, ipeirotis2010demographics, ross2010crowdworkers}. This country distribution is likely due to the fact that 
workers can receive money from AMT in the USA and in India while workers from other countries are paid in Amazon.com gift cards \cite{amtpay}. 

Studies that investigated crowdworker motivation suggest that there are different motivations for participating on micro-task platforms. For example, one early study on the reasons crowdworkers have for participating on AMT was conducted by Ipeirotis \cite{ipeirotis2010demographics}. In this study, the author asked the multiple-choice question ``Why do you complete tasks in Mechanical Turk?'', offering six response options. He found that more Indians than Americans treat AMT as a primary source of income, and that few Indian workers report the reason ``To kill time.''
Hossain \cite{hossain2012users} created a classification of motivation in online platform participation, listing extrinsic and intrinsic motivators and incentives.

Kaufmann et al. \cite{kaufmann2011more} developed an early model for measuring crowdworker motivations on AMT, differentiating between enjoyment based motivation, community based motivation, immediate payoffs, delayed payoffs and social motivation.
They used a sample composed of Indian and US workers on AMT and found that the construct with the highest score was ``immediate payoffs,'' i.e. payment. 
Their study further found that the pastime score correlated positively with household income and negatively with the weekly time spent on AMT, and that workers who spend a lot of time on AMT may be motivated differently than workers who spend little time on the platform.

Antin and Shaw \cite{antin2012social} used a list experiment to investigate social desirability effects in motivation self-reports of crowdworkers from the USA and India on AMT. Using the four items ``to kill time,'' ``to make extra money,'' ``for fun'' and ``because it gives me a sense of purpose,'' they found that US workers tended to over-report all four reasons while Indian workers tended to over-report ``sense of purpose'' and under-report ``killing time'' and ``fun.''

For measuring extrinsic motivations of crowdworkers, Naderi et al. \cite{naderi2014development} evaluated a 4-factor model using a subset of WEIMS items on a sample of US workers on AMT. In this model, identified and integrated regulation are merged into one factor, and the intrinsic motivation factor is omitted. After a first version of the present study was published \cite{mcms17}, Naderi adapted and extended their scale to include intrinsic motivation \cite{naderi2018measure}. The adapted scale, named the Crowdwork Motivation Scale, measures five constructs, three of which are measured by WEIMS items (amotivation, external regulation and identified regulation). 
The scale was evaluated on three samples of workers on AMT (N = 170, 90 and 86).

In addition to these few quantitative studies, several qualitative studies on the motivations of crowdworkers have been conducted. For example, Gupta et al. \cite{gupta2014understanding, gupta2014turk} investigated, among other aspects, the motivations of Indian crowdworkers on AMT and Martin et al. \cite{martin2014being} studied the content of a forum for AMT users.
Other research related to the motivations of crowdworkers includes measuring the impact of motivation on performance \cite{rogstadius2011assessment} and manipulating motivations via task framing \cite{chandler2013breaking} or achievement feedback \cite{lee2013experiments}.

In sum, these studies demonstrate that there are meaningful differences in crowdworkers' motivations to participate on micro-task platforms. However, systematic and theory-driven inquiries into the motivations of crowdworkers remain in short supply. SDT-based work motivation scales may offer a suitable foundation for such inquiries, but the applicability of such scales to the micro-task domain has yet to be established.

\section{Suitability of Existing Work Motivation Scales}
\label{sec:existing}

Crowd employment on micro-task platforms is similar to traditional employment in the sense that in both contexts, workers provide a service in exchange for payment. In both contexts, tasks need to be completed, and these tasks are usually specified by the employer/requester and executed by the employee/crowdworker. However, several fundamental aspects of work on micro-task platforms differ from the traditional employment context. For example, on micro-task platforms, the relationship between the requester and the worker is often completely anonymous and extremely short-lived, often lasting only a few minutes. Furthermore, there is only a minimal amount of communication between the requester and the workers, often not exceeding the static one-way communication via written task instructions. Regarding the social environment, there is often no communication or collaboration between co-crowdworkers, as micro tasks are intended to be completed as individual work.

In order to determine to what extent existing SDT-based work motivation scales that were developed for the traditional work context are suitable for application in the micro-task context, we conducted an evaluation of two work motivation scales, the WEIMS \cite{weims} and the MWMS \cite{mwms}, with crowdworkers on CrowdFlower.\footnote{CrowdFlower changed its name to Figure Eight in February 2018 \cite{figureeightarticle}.} The micro-task market is dominated by two platforms, CrowdFlower and AMT, which are estimated to share 80\% of all revenue generated in the micro-task market, with revenues being approximately equal \cite{kuek2015global}.
Our reason for choosing CrowdFlower over AMT is that we aim to provide a motivation scale suitable for an international comparison of crowdworker motivations, instead of exclusively focusing on crowdworkers based in the USA and in India. We consider CrowdFlower to be better suited for this task as it pays workers via independent partner channels\footnote{http://www.crowdflower.com/labor-channels/} and therefore attracts a more international crowd-workforce.  

In order for the scale stems and items of the WEIMS and the MWMS to be conceptually applicable to the crowdworking domain, we had to make minimal adaptations to the scales before evaluating them in the micro-task context. For WEIMS, we changed the stem ``Why do you do your work?'' to ``Why do you do CrowdFlower tasks?''\footnote{We chose to use the term ``CrowdFlower tasks'' in the stem questions and in the items instead of a more general term because workers who are logged into CrowdFlower via the partner channels see that they are doing ``CrowdFlower tasks.'' We can therefore assume that workers know what CrowdFlower tasks are. In contrast, general terms like ``micro tasks'' are widely used in scientific publications and sometimes in the media but we cannot be sure that workers on CrowdFlower understand this term as it does not appear frequently on partner channel websites or on the platform itself.} and replaced the word ``it'' (referring to ``your work'') in the items  with ``CrowdFlower tasks.''
The stem of MWMS ``Why do you or would you put efforts into your current job?'' was changed to ``Why do you or would you put efforts into CrowdFlower tasks?'' 
and words in the items referring to ``your current job'' were replaced with ``CrowdFlower tasks.'' Additionally, one item in the MWMS was conceptually not applicable to the domain and had to be adapted. There is no equivalent to ``losing one's job'' on micro-task platforms. The closest concept on CrowdFlower is failing many quality control questions, which results in a lower worker account accuracy and consequently in less tasks being offered to the worker. Therefore, the item ``Because I risk losing my job if I don't put enough effort in it.'' was changed to ``Because I risk not being offered enough tasks if I don't put enough effort into them.''.

Respondents answered both scales along a 7-point Likert-type scale. We adopted the verbal descriptions of the scale's endpoints from the original scales: For the adapted WEIMS (A-WEIMS), the scale ranged from ``does not correspond at all'' (1) to ``corresponds exactly'' (7) and for the adapted MWMS (A-MWMS), the scale ranged from ``not at all'' (1) to ``completely'' (7).

For both the A-WEIMS and the A-MWMS, we collected responses from 500 crowdworkers residing in the USA. The surveys were posted as tasks on CrowdFlower and  anonymity was ensured in the description of the tasks.
After removing spammers (also see Section \ref{sec:validation}),
the sample size was 424 for the A-WEIMS and 414 for the A-MWMS. This entails a subject-to-item ratio higher than 20:1, which is a suitable ratio for factor analysis \cite{osborne2009best, froman2001elements}. 
The questionnaires adhered to the default design of the CrowdFlower platform and the design was very similar to that used in the later validation of the MCMS scale (see figures in Appendix~\ref{app:interface}).

\textbf{Confirmatory Factor Analysis.} 
Confirmatory factor analysis (CFA) is a multivariate data analysis technique used to test how well latent constructs, specified according to theory, represent reality according to the data gathered \cite{hair2018multivariate}.
We used CFA to evaluate the psychometric quality (i.e., the quality of the measurement of the latent constructs) of the A-WEIMS and the A-MWMS. In particular, we evaluated the factor structure of the items with CFAs, using the R packages lavaan \cite{lavaan} and semtools \cite{semtools}.

Because of non-normality of the item distributions, we used a robust maximum likelihood estimator (as suggested in e.g. \cite{finney2006non, de2013construct}). By specifying a robust maximum likelihood estimator, the model parameters were estimated with robust standard errors, and a Satorra-Bentler (S-B) scaled test statistic is reported \cite{satorra2010ensuring, lavaan}. 

\begin{table}[b!]
\centering
\caption{\textbf{Evaluation of existing work motivation scales.} This table shows the goodness-of-fit measures for the different models based on existing SDT-based work motivation scales which were minimally adapted to the crowdworking domain.}
\label{table:cfa_existing}
%\resizebox{\columnwidth}{!}{
\begin{tabular}{lcccccccc}
\toprule
\textbf{Scale/Model} & \textbf{N} & \textbf{S-B}$\bm{\chi^2}$ & \textbf{df} & \textbf{CFI} & \textbf{TLI} & \textbf{RMSEA} & \textbf{RMSEA 90\% CI} & \textbf{SRMR} \\ \midrule
\texttt{A-WEIMS-M2}  &  424 &  500.58	& 125 &	0.908 & 0.888 &	0.084 & 0.077 \; 0.091 & 0.067\\  
\texttt{A-WEIMS-M3}  &  424 &  110.57	& 48 &	0.969 & 0.957 &	0.055 & 0.043 \; 0.068 & 0.043\\
%\midrule
\texttt{A-MWMS-M1} & 414 &  828.27 &  143   & 0.818 & 0.782  & 0.108  & 0.101 \; 0.114 & 0.170\\ 
\texttt{A-MWMS-M3} & 414 &   667.90   & 139 & 0.859 & 0.827 &  0.096  & 0.089 \; 0.103 & 0.155 \\ 
\texttt{A-MWMS-M4} & 414 &   532.55   & 137 & 0.895 & 0.869 &  0.084  & 0.077 \; 0.091 & 0.087 \\ \bottomrule
\end{tabular}
%}
\end{table}

In the context of CFA, the validity of a measurement model is evaluated via a range of goodness-of-fit (GOF) measures. These GOF measures assess the extent to which the theory (as specified in the model) represents reality (i.e., the data). Once acceptable levels of GOF measures are established, other aspects of construct validity can be evaluated \cite{hair2018multivariate}. For a further discussion of construct validity, see Section~\ref{sec:validation}.
We evaluated the model fit based on the absolute\footnote{Absolute GOF measures measure how well the model represents the data, independently of other, alternative, models \cite{hair2018multivariate}.} fit measures \emph{root mean squared error of approximation} (RMSEA) and \emph{standardised root mean square residual} (SRMR) as well as the incremental\footnote{Incremental GOF measures compare the specified model with a baseline model where all variables are uncorrelated \cite{hair2018multivariate}.} fit measures \emph{comparative fit index} (CFI) and \emph{Tucker-Lewis index} (TLI). 
In line with current conventions for judging model fit (e.g. \cite{kline2015principles, marsh2005goodness}), we chiefly relied on the CFI, RMSEA and and SRMR to assess model fit\footnote{A discussion on the guidelines for determining model fit can be found in Hooper et. al \cite{hooper2008structural}.} and judged model fit to be acceptable according to the following criteria: A well-fitting model should have an RMSEA of less than 0.06, a SRMR of less than 0.08, and a CFI and TLI higher than 0.95 (but at least 0.9 to be acceptable).
Furthermore, we report the (S-B scaled) Chi-Square test statistic but do not rely on it for determining model fit as it is very sensitive to sample size (e.g. \cite{bentler1980significance}).

\textbf{Measurement Models.}
We tested the following models for the adapted WEIMS: (1) The original WEIMS model with six factors (\texttt{A-WEIMS-M1}); (2) an alternative five-factor WEIMS model with identified regulation and integrated regulation loading onto a single factor (\texttt{A-WEIMS-M2}); and (3) the four-factor, 12-item subset of WEIMS items used by Naderi et al. \cite{naderi2014development} to measure the extrinsic motivations of workers on Amazon Mechanical Turk (\texttt{A-WEIMS-M3}).\footnote{We did not include the extended and adapted version of the 12-item subset of WEIMS items \cite{naderi2018measure} in the evaluation as it was published a year after the development and validation of the MCMS was completed and made available as a first version \cite{mcms17}.} 
Our rationale for evaluating the alternative model \texttt{A-WEIMS-M2} is that the integrated regulation factor has been shown to be poorly separable from identified regulation and intrinsic motivation (e.g. \cite{weims, vallerand1992academic}), which is also one of the reasons for why the MWMS does not include an integrated regulation factor \cite{mwms}.

For the A-MWMS, we tested the model originally hypothesized by Gagné et al. \cite{mwms} (\texttt{A-MWMS-M1}) and the model which had the best fit in \cite{mwms} (\texttt{A-MWMS-M2}). Furthermore, we tested a six-factor model in which material external and social regulation are separate factors, omitting the second-order external regulation factor. We tested this model with a hypothesized correlation of zero between intrinsic motivation and both external regulation factors (\texttt{A-MWMS-M3}) as well as without this correlation restriction (\texttt{A-MWMS-M4}).

\textbf{Results.}
Table \ref{table:cfa_existing} shows the goodness-of-fit statistics of the different models. None of the evaluated models, with the exception of the four-factor model \texttt{A-WEIMS-M3}, which does not measure intrinsic motivation or social external regulation, had an acceptable model fit on our data.

The A-WEIMS model with six factors (\texttt{A-WEIMS-M1}) could not be estimated due to the factors identified regulation and integrated regulation not being distinguishable from another, which resulted in the covariance matrix of the factors not being positive definite. This is consistent with the findings of Naderi et al. \cite{naderi2014development}. Also for the alternative five-factor model \texttt{A-WEIMS-M2}, the goodness-of-fit measures were outside the acceptable range.

The four-factor model \texttt{A-WEIMS-M3} was the only evaluated model with a good fit.\footnote{Note, however, that this evaluated subset of A-WEIMS is not identical to the one used by Naderi et al. \cite{naderi2014development} as a result of the adaptations made in item wording. Naderi et al. \cite{naderi2014development} reported a lower CFI (0.931) and a higher RMSEA (0.088).} However, besides the drawbacks that the model does not measure intrinsic motivation or social external regulation, and two of the four factors are measured by only two items each, it has additional limitations: It includes items for measuring the amotivation construct which were criticized by Gagné et al. \cite{mwms} for resembling low satisfaction of the need for competence rather than measuring amotivation (e.g., ``I ask myself this question, I don't seem to be able to manage the important tasks related to this work.''). Besides this criticism regarding content validity (i.e., regarding the extent to which the items correspond to the conceptual definition of the construct \cite{hair2018multivariate}), the amotivation construct also had an average variance extracted (AVE) below $0.5$ (also see Section~\ref{sec:validation}), suggesting low convergent validity \cite{hair2018multivariate}.

An examination of the estimated correlations of \texttt{A-WEIMS-M3} reveals that the second-strongest positive correlation (0.38) is between the amotivation construct and the identified/integrated regulation construct.\footnote{Naderi et al. \cite{naderi2014development} reported an even higher correlation of 0.48.} This questions nomological validity (i.e., whether the relationships between the constructs correspond to theory and prior research \cite{hair2018multivariate}) as it is inconsistent with both SDT and the correlation patterns reported by other SDT-based motivation scales, which report a low to moderate negative correlation between these constructs (e.g. \cite{weims, mwms, walking, teaching}).

Of the evaluated A-MWMS models, the six-factor model with separate factors for social external and material external regulation and no correlation restrictions had the best fit. However, the fit was still not acceptable, with all goodness-of-fit measures falling outside acceptable ranges. 
The fit measures for \texttt{A-MWMS-M2} are not included in Table \ref{table:cfa_existing} because the covariance matrix of the factors was not positive definite. In sum, these results suggest that the existing scales do not allow to measure the crowdworker motivation validly. We thus decided to conduct an in-depth item analysis.

In order to identify which items in A-WEIMS and A-MWMS were most problematic in the micro-task context, we conducted an exploratory factor analysis (EFA) on both samples. In contrast to CFA, factors in EFA are not specified according to theory, but derived from the data \cite{hair2018multivariate}. Therefore, this technique is useful for investigating the underlying structure of a set of items, using the gathered data as a starting point. For all exploratory factor analyses, we used oblique rotation (promax) because – in line with SDT – we expected the factors (i.e., motivational dimensions) to correlate.

For the A-WEIMS, we conducted an EFA with four factors: We removed the amotivation items due to the problems described above and, because of the CFA results, we expected the integrated regulation items and the identified regulation items to load onto a single factor.\footnote{An EFA with five factors showed that one item from the identified regulation construct and one item from the integrated regulation construct loaded onto a separate factor, while all other items from these constructs loaded onto a single factor. Therefore, we proceeded with the four-factor version.} The most problematic remaining items were ``Because this type of work provides me with security'', which had the highest loading on the the same factor as the items from the identified/integrated regulation factor and ``Because I want to be a `winner' in life'', which did not have a high loading on any factor. This item was also critiziced by Gagn\'e et al. \cite{mwms} for being culturally sensitive.
Removing these items yielded a solution with four factors (material external regulation, introjected regulation, identified/integrated regulation and intrinsic motivation), whereby two of the factors only had two items remaining.

For the A-MWMS, we conducted an EFA with six factors.\footnote{An EFA with five factors showed that the items from the material and social external regulation constructs did not load onto the same factor.} Again, one of the most problematic items was related to security (``Because it gives me greater job security if I put enough efforts into doing CrowdFlower tasks.''), which did not load onto one factor with the other items from the material external regulation factor, but instead had a high loading on the same factor as the items from the identified regulation construct. Also two items from the introjected regulation construct (``Because I have to prove to myself that I can.'' and ``Because it makes me feel proud of myself.'') had moderately high loadings (0.57 and 0.58) on this factor.
After removing these three items, the introjected regulation construct as well as several other items were still problematic. Iteratively removing all problematic items yielded a solution with amotivation measured by two items, external material regulation measured by only one item, and either no introjected regulation factor or one that was very poorly distinguishable from the social external regulation factor.

Thus, our results show that the evaluated work motivation scales developed for the traditional work context do not work well within the crowdworking context when only minimal adaptations in item wording are made. While both scales contain items that are potentially useful for measuring the motivations of  crowdworkers, neither of the scales is an accurate measure of the full spectrum of motivational dimensions in the micro-task context. For the development of a reliable motivation scale that fills these gaps and measures the motivations of crowdworkers on all dimensions proposed by SDT, further adaptations are needed.

\section{Development of the MCMS}
\label{sec:development}

The results of our factor analyses conducted on the slightly modified WEIMS and MWMS underscore the necessity for developing a new scale for measuring the motivations of crowdworkers that is adapted to the idiosyncracies of the crowdwork environment.
To meet this necessity, we developed the Multidimensional Crowdworker Motivation Scale (MCMS). 
The MCMS was developed to provide a psychometrically sound scale that covers the motivational dimensions proposed by self-determination theory and that can be answered by crowdworkers in a limited amount of time.

We proceeded in three steps. First, we compiled a pool of items conceptually suitable for the characteristics of the crowdworking domain. An item pool is a set of candidate items that reflect the latent constructs the scale intends to measure. During the development of a scale, this set of candidate items is then reduced and refined by deleting items that exhibit undesirable properties such as high loadings on multiple factors or no high loadings on any factor, with the goal of arriving at a final scale with optimized reliability and scale length (e.g. \cite{devellis2016scale, worthington2006scale}). 
Second, we thus selected items from the pool based on an exploratory factor analysis on a sample of workers residing in the USA. Third, we further reduced and refined the item pool based on exploratory factor analyses on samples from Spanish\footnote{In this paper, we use country demonyms synonymously with the location of workers for better readability.} and Indian crowdworkers.

\textbf{Item Pool Generation.} The results of our evaluation described in the previous section indicate that both the A-WEIMS and the A-MWMS contain items that are potentially useful for application in the micro-task context.
Therefore, for compiling the item pool, we first included all items from the A-WEIMS\footnote{Notably, this also included all items from \texttt{A-WEIMS-M3} except for the amotivation items due to the problems described in Section \ref{sec:existing}.} and the A-MWMS that were not among the most problematic items according to the results of the factor analyses. We retained the adaptations to item wording.
To extend the pool, we added semantically suitable items from the SDT-based motivation scales developed in \cite{walking} and \cite{teaching} as well as nine new items developed by the authors.\footnote{The development of the new items focused mainly on the material external regulation construct, as material external regulation is a construct present only in scales intended to measure motivation in a context where material rewards (such as money) are relevant. It is therefore not present in most SDT-based motivation scales.} The total number of items in the pool was 44. 

To ensure content validity (i.e., the extent to which the items correspond to the theoretical constructs \cite{hair2018multivariate}), we based the majority of candidate items on existing scales and we closely followed the definitions of the constructs during the selection of candidate items from existing scales as well as during the creation of the new items. We used the construct definitions provided in the Handbook of Self-Determination Theory Research \cite{deci2002handbook} and a publication by Gagné and Deci discussing the constructs in the context of work motivation \cite{gagne2005self}.

We used the stem phrasing ``Why do you or would you put efforts into doing CrowdFlower tasks?'', adapted from MWMS, in order to capture both actual and latent motivations.  The items were rated along a 7-point Likert-type scale ranging from ``not at all'' (1) to ``completely'' (7).\footnote{The verbal descriptions for each scale point were adopted from the MWMS \cite{mwms} and shown to the participants in the task instructions: 1 = ``not at all'' , 2 = ``very little'', 3 = ``a little'', 4 = ``moderately'', 5 = ``strongly'', 6 = ``very strongly'', 7 = ``completely''.}

\textbf{First Round of Data Collection.} For a first selection of items from the pool, we collected answers from 1,000 crowdworkers residing in the USA and conducted an exploratory factor analysis on their responses. 
Consistent with the findings from Section~\ref{sec:existing}, we found that the items intended to measure material external regulation and social external regulation did not load on the same factor.
Therefore, we aimed at a six-factor model with the two separate external regulation factors (social and material).
Items that had insufficient loadings on the appropriate factor ($<$ 0.5), items which loaded on a factor other than the hypothesized one, and items with high cross-loadings ($>$ 0.35) were iteratively removed from the initial pool, creating a reduced item pool with 36 items.

\textbf{Second Round of Data Collection.} We conducted a second round of data collection with this reduced item pool, collecting responses from 1,000 Spanish and 1,200 Indian crowdworkers.\footnote{India was selected because it has a significant population of crowdworkers on different platforms. Spain was selected in order to include a European country with a significant population of crowdworkers.} The additional 200 responses from the Indian crowdworkers were requested because of the high amount of spam received in this group (also see Section \ref{sec:validation}), with the aim of achieving an item-to-response ratio of close to 1:20.
Again, we iteratively removed items with low loadings (with the higher threshold of $<$ 0.7 if more than three items were left for this construct) or high cross-loadings (with a threshold of $>$ 0.3). Furthermore, if two items were phrased very similarly and the factor had more than three items remaining, we removed the item with the lower loading.

The final MCMS contains 18 items, with three items measuring each factor. Of the 18 final items, five items (Am2, Introj2, Introj3, Intrin1, Intrin3) were adapted from \cite{mwms}, four items (ExMat2, Ident1, Ident2, Ident3) from \cite{weims}, two items (Am3, Introj1) from \cite{teaching}, two items (ExSoc2, Intrin2) from \cite{walking} and five items (Am1, ExMat1, ExMat3, ExSoc1, ExSoc3) are new (but semantically based on items from existing scales). Like other SDT-based work motivation scales such as the MWMS \cite{mwms}, the MCMS aims to measure motivations for putting effort into the job (in this case micro tasks) in general, as opposed to measuring the motivations for specific tasks within a job. Table~\ref{table:mcms_items} in Appendix~\ref{app:scale} shows the scale.

\section{Validation of the MCMS}
\label{sec:validation}

\textbf{Data Collection.} 
With the final 18-item version of the MCMS, we collected data from 10 countries, with 900 participants from each country, for validation. 
We selected countries from three World Bank income groups\footnote{http://databank.worldbank.org/data/download/site-content/CLASS.xls}: high income, upper middle income and lower middle income. From each of the three income groups, we selected three countries with high activity on CrowdFlower.
The countries were selected according to the following criteria: First, the country had to be active on CrowdFlower (either high in the Alexa\footnote{http://www.alexa.com/} ranking or one of the top contributing countries in at least one of the partner channels). Second, we aimed for a high cultural diversity overall as well as within the income groups.
For the high income group, the selected countries were USA, Germany and Spain. The upper middle income group contains Brazil, Russia and Mexico, and the lower middle income group is comprised of India, Indonesia and the Philippines. 
Note that in the rest of this paper, we use the group label ``Middle Income'' (MID) for the upper middle income group and ``Low Income'' (LOW) for the lower middle income group for better readability.

In addition, we collected responses from Venezuela because it was the most active country on CrowdFlower at the time of data collection, with CrowdFlower receiving 18.5\% of traffic from this country.\footnote{Data obtained from http://www.alexa.com/.} However, we did not include Venezuela in the data grouped by income because we believe it represents a special case:  At the time of data collection, the US\$ earned on CrowdFlower could be sold on the black market at a rate several orders of magnitude higher than the official exchange rate \cite{blackmarket}.

In order to capture a diverse sample of crowdworkers in each country, the starting times of the survey were divided into three groups: (1) 300 responses were requested during typical working hours (8:00 am to 5:00 pm in the appropriate time zone), (2) 300 responses were requested in the evening (6:00 pm to 11:00 pm in the appropriate time zone) and finally, (3) 300 responses were requested during weekends. We made the survey available to workers of all CrowdFlower levels. The data was collected in October and November 2016. A full description of the demographic characteristics of our sample can be found in Posch et al.  \cite{posch2018characterizing}.

\begin{table}[h!b]
\caption{\textbf{Sample sizes and percentage of spam received.} This table shows the sample sizes of the different groups before and after spam removal, as well as the percentage of spam received.}
\centering
%\resizebox{.55\textwidth}{!}{
\begin{tabular}{l l c c c} 
\toprule
\textbf{Group} & \textbf{Code} & $\mathbf{N}_{raw}$ & \textbf{Spam} & $\mathbf{N}_{clean}$  \\  
 \midrule
 All & ALL & 9000 & 35 \% & 5857  \\
 
 \midrule
 
 High Income & HIGH & 2700 & 28 \% & 1952 \\
 Middle Income & MID & 2700 & 32 \% & 1834 \\
 Low Income & LOW & 2700 & 44 \% & 1508 \\
 \midrule

 USA & USA & 900 & 20 \% & 721\\
 Spain & ESP & 900 & 25 \% & 677  \\
 Germany & DEU & 900 & 38 \% & 554 \\%[1mm]
 
 Brazil & BRA & 900 & 45 \% & 496 \\
 Russia & RUS & 900 & 25 \% & 677\\%[1mm]
 Mexico & MEX & 900 & 27 \%  & 661\\

 India & IND & 900 & 32 \% & 608 \\
 Indonesia & IDN & 900 & 55 \% & 401  \\
 Philippines & PHL & 900 & 45 \% & 499  \\%[1mm]
 
 Venezuela & VEN & 900 & 37 \% & 563 \\
 \bottomrule
\end{tabular}
%}
\label{table:demographic}
\end{table}

\textbf{Task Interface and Payment.} The items in the MCMS were randomly permuted and presented to crowdworkers as a task on CrowdFlower. Besides the MCMS items, the task also included a section with demographic questions and questions about money use, as well as a section in which workers were instructed to think of five reasons for why they do tasks on CrowdFlower and asked to write down these reasons. Anonymity was guaranteed in the task instructions. The interface\footnote{English was chosen as the interface language for all countries for two reasons: Firstly, CrowdFlower's default interface language is English and all workers are expected by the platform to understand instructions in English. This is underscored by the fact that ``English'' is not selectable in the requester interface when choosing crowdworkers of a specific language. Secondly, translating stems and items has a risk of introducing semantic mismatches. We hence weighted the risk of introducing translation mismatches higher than potential errors due to non-native speakers' misinterpretations.} of the task is shown in Appendix \ref{app:interface}. After completing a task, crowdworkers on CrowdFlower are asked by the platform to judge the task according to different criteria, one of them being the clarity of the task instructions and interface. The question asked is \emph{``How clear were the task instructions and interface?''} and workers are asked to answer on a five-point scale ranging from ``very unclear'' to ``very clear.'' The average of the workers' responses was higher than 4.0 in all countries except for Brazil (3.6) and Indonesia (3.8). This indicates that the task instructions and interface was perceived as clear in most countries, and as ``somewhat clear'' in Brazil and Indonesia.

We aimed for a payment similar to most other tasks on CrowdFlower in order to minimize population bias in the responses. Based on studies reporting average earnings on micro-task platforms (e.g. \cite{berg2016income, horton2010labor, khanna2010evaluating}) and on the first author’s experience as a crowdworker on CrowdFlower, we payed US\$0.1 for the task excluding platform fees. The survey presented by the platform to workers after completing a task includes a question about task payment (\emph{``How would you rate the pay for this task relative to other tasks you've completed?''}) that crowdworkers answer on a five-point scale ranging from ``much worse'' to ``much better''. For the different countries, the averages of the workers' responses ranged from 3.5 (in Germany) to 4.1 (in Mexico and India), indicating that our task payment was equal to or ``somewhat better'' than other tasks according to the workers' perception.

\textbf{Spam Detection.} 
We expected a significant amount of spam in the responses, such as people not reading the questions and clicking randomly or workers accepting the task despite having insufficient English skills. To counteract a high amount of noise in the dataset, we included three test items in the motivation scale section of the CrowdFlower task, and an additional test question in the demographics section.\footnote{As CrowdFlower does not offer built-in quality control mechanisms for survey-type tasks, we did not use any platform-specific quality control mechanisms.} Employing test questions is a common method to implement attention checks in survey design (e.g., \cite{krosnick1999survey, oppenheimer2009instructional, vannette2014comparison}) and has also been employed in crowdsourcing research tasks (e.g., \cite{paolacci2010running}).

The three test items in the motivation scale section of the task asked participants to answer with a specific ranking on a 7-point scale, and the test question in the demographics section consisted of the question ``Are you paying attention to the questions?'' with the possible answers ``No,'' ``Yes'' and ``I don't know'' selectable from a drop-down list. These questions ensured that less than 0.1\%  ($(\tfrac{1}{7})^3*\tfrac{1}{3}$) of spammers passed the test questions, assuming that all four questions were answered at random.
Table \ref{table:demographic} shows the percentage of spam and the sample size after spam removal in each country and income group. The table also introduces the country and group codes used in the remainder of this paper.

\begin{figure}[t!]
\centering
    \scalebox{.8}{\begin{tikzpicture}[auto,node distance=.20cm,
    latent/.style={ellipse,draw,very thick,inner sep=0pt,minimum height =13mm, minimum width = 35mm,align=center, font=\fontsize{10pt}{8pt}\selectfont},
    manifest/.style={rectangle,draw,very thick,inner sep=0pt,minimum width=17mm,minimum height=5mm},
    paths/.style={->, very thick, >=stealth'},
    twopaths/.style={<->, very thick, >=stealth'}
]

% Define observed variables
\node [manifest] (Am1) at (0,0) {Am1};
\node [manifest] (Am2) [below=of Am1]  {Am2};
\node [manifest] (Am3) [below=of Am2]  {Am3};

\node [manifest] (ExMat1) [below=.3cm of Am3]  {ExMat1};
\node [manifest] (ExMat2) [below=of ExMat1]  {ExMat2};
\node [manifest] (ExMat3) [below=of ExMat2]  {ExMat3};

\node [manifest] (ExSoc1) [below=.3cm of ExMat3]  {ExSoc1};
\node [manifest] (ExSoc2) [below=of ExSoc1]  {ExSoc2};
\node [manifest] (ExSoc3) [below=of ExSoc2]  {ExSoc3};

\node [manifest] (Introj1) [below=.3cm of ExSoc3]  {Introj1};
\node [manifest] (Introj2) [below=of Introj1]  {Introj2};
\node [manifest] (Introj3) [below=of Introj2]  {Introj3};

\node [manifest] (Ident1) [below=.3cm of Introj3]  {Ident1};
\node [manifest] (Ident2) [below=of Ident1]  {Ident2};
\node [manifest] (Ident3) [below=of Ident2]  {Ident3};

\node [manifest] (Intrin1) [below=.3cm of Ident3]  {Intrin1};
\node [manifest] (Intrin2) [below=of Intrin1]  {Intrin2};
\node [manifest] (Intrin3) [below=of Intrin2]  {Intrin3};

% Define latent variables
\node [latent] (AMOTIVATION) [right=1.0cm of Am2] {Amotivation};
\node [latent] (EXTERNAL_MAT) [right=1.0cm of ExMat2] {Material Ext.\\ Regulation};
\node [latent] (EXTERNAL_SOC) [right=1.0cm of ExSoc2] {Social Ext.\\ Regulation};
\node [latent] (INTROJECTED) [right=1.0cm of Introj2] {Introjected\\ Regulation};
\node [latent] (IDENTIFIED) [right=1.0cm of Ident2] {Identified\\ Regulation};
\node [latent] (INTRINSIC) [right=1.0cm of Intrin2] {Intrinsic\\ Motivation};

% Draw paths from latent to observed variables
\foreach \all in {Am1, Am2, Am3}{
    \draw [paths] (AMOTIVATION) to node { } (\all);
}

\foreach \all in {ExMat1, ExMat2, ExMat3}{
    \draw [paths] (EXTERNAL_MAT) to node { } (\all);
}

\foreach \all in {ExSoc1, ExSoc2, ExSoc3}{
    \draw [paths] (EXTERNAL_SOC) to node { } (\all);
}

\foreach \all in {Introj1, Introj2, Introj3}{
    \draw [paths] (INTROJECTED) to node { } (\all);
}

\foreach \all in {Ident1, Ident2, Ident3}{
    \draw [paths] (IDENTIFIED) to node { } (\all);
}

\foreach \all in {Intrin1, Intrin2, Intrin3}{
    \draw [paths] (INTRINSIC) to node { } (\all);
}

\draw [twopaths] (AMOTIVATION.east) to [bend left=90] (EXTERNAL_MAT.east);
\draw [twopaths] (AMOTIVATION.east) to [bend left=90] (EXTERNAL_SOC.east);
\draw [twopaths] (AMOTIVATION.east) to [bend left=90] (INTROJECTED.east);
\draw [twopaths] (AMOTIVATION.east) to [bend left=90] (IDENTIFIED.east);
\draw [twopaths] (AMOTIVATION.east) to [bend left=90] (INTRINSIC.east);

\draw [twopaths] (EXTERNAL_MAT.east) to [bend left=90] (EXTERNAL_SOC.east);
\draw [twopaths] (EXTERNAL_MAT.east) to [bend left=90] (INTROJECTED.east);
\draw [twopaths] (EXTERNAL_MAT.east) to [bend left=90] (IDENTIFIED.east);
\draw [twopaths] (EXTERNAL_MAT.east) to [bend left=90] (INTRINSIC.east);

\draw [twopaths] (EXTERNAL_SOC.east) to [bend left=90] (INTRINSIC.east);
\draw [twopaths] (EXTERNAL_SOC.east) to [bend left=90] (INTROJECTED.east);
\draw [twopaths] (EXTERNAL_SOC.east) to [bend left=90] (IDENTIFIED.east);

\draw [twopaths] (INTROJECTED.east) to [bend left=90] (IDENTIFIED.east);
\draw [twopaths] (INTROJECTED.east) to [bend left=90] (INTRINSIC.east);

\draw [twopaths] (IDENTIFIED.east) to [bend left=90] (INTRINSIC.east);

\end{tikzpicture}}
    \caption{\textbf{Factor structure of the MCMS.} This figure shows the factor structure of the hypothesized MCMS model. Items (i.e. measured variables) are represented by rectangles, ovals represent latent constructs, curved arrows indicate a correlational relationship between latent constructs and straight arrows indicate the dependence relationship between latent constructs and items. The individual items are shown in Appendix~\ref{app:scale}.} 
    \label{fig:sem_diagram}
\end{figure}
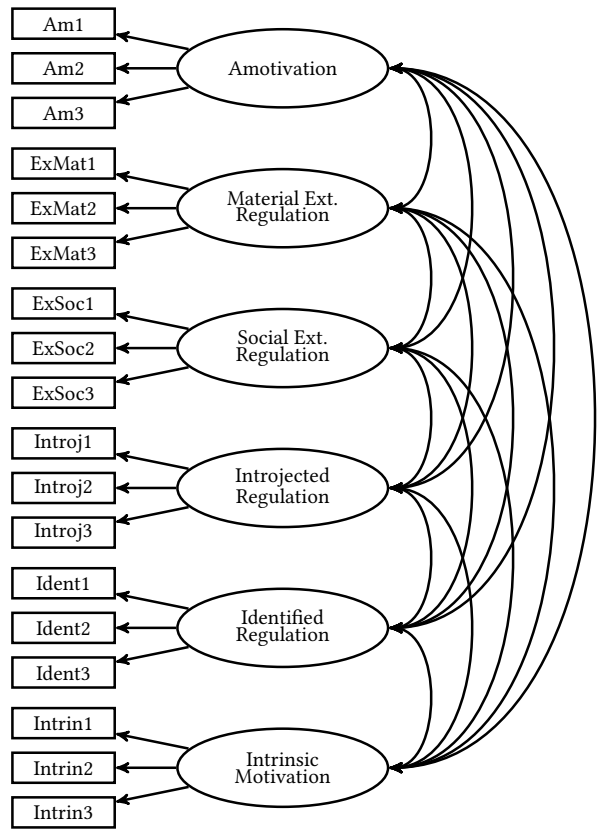

\textbf{Hypothesized Model.}
Our hypothesized model measures six constructs and is depicted in Figure~\ref{fig:sem_diagram}. The inclusion of a social external regulation construct in addition to material external regulation was adopted from the MWMS \cite{mwms} because both social and material rewards are important in the work context \cite{mwms, stajkovic1997meta}. As suggested by our evaluation of existing scales and our results from exploratory factor analysis on the MCMS item pool, we modeled material external and social external regulation as two separate factors. 
Hence, our hypothesized model is a six-factor model in which all factors are first-order factors. Note that the material and social external regulation are not adjacent factors in the continuum hypothesized by SDT but occupy the same spot.

Whereas SDT hypothesizes that intrinsic motivation does not correlate with external regulation \cite{mwms, ryan1989perceived}, 
Chemolli and Gagné \cite{chemolli2014evidence} found that intrinsic motivation correlates with external regulation significantly for both the MWMS and the Academic Motivation Scale \cite{vallerand1992academic}. 
Based on these findings, we decided not to restrict these correlations to zero in our hypothesized model but to evaluate the model fit of both the correlation-restricted and the unrestricted model. No cross-loadings were hypothesized.

\textbf{Descriptive Statistics and Internal Consistency.} 
Table~\ref{table:means_sds} summarizes the observed factor means as well as standard deviations for each of the countries and income groups in our data. Table~\ref{table:factor_correlations} displays the Pearson correlations of the observed factor means. 
\begin{table}[b!]
\caption{\textbf{Manifest scale means and standard deviations.} This table shows the manifest scale means and standard deviations for all groups and factors.}
\centering
%\resizebox{\columnwidth}{!}{
\begin{tabular}{l c c c c c c} 
\toprule
  \textbf{Group} & \textbf{Amotivation} & \textbf{Material} & \textbf{Social} & \textbf{Introjected} & \textbf{Identified} & \textbf{Intrinsic}\\
\midrule
ALL & \msdd{1.84}{1.09} & \msdd{6.05}{1.08} & \msdd{2.47}{1.58} & \msdd{2.25}{1.46} & \msdd{4.27}{1.73} & \msdd{5.67}{1.26}\\ \midrule
HIGH & \msdd{2.06}{1.21} & \msdd{5.86}{1.18} & \msdd{1.99}{1.37} & \msdd{1.91}{1.29} & \msdd{3.56}{1.73} & \msdd{5.27}{1.37} \\
MID & \msdd{1.86}{1.06} & \msdd{6.12}{1.02} & \msdd{2.68}{1.60} & \msdd{2.59}{1.58} & \msdd{4.53}{1.65} & \msdd{5.82}{1.18} \\
LOW & \msdd{1.72}{1.02} & \msdd{6.13}{0.99} & \msdd{2.74}{1.68} & \msdd{2.31}{1.44} & \msdd{4.60}{1.57} & \msdd{5.90}{1.11} \\
\midrule
USA & \msdd{1.91}{1.17} & \msdd{5.86}{1.25} & \msdd{1.67}{1.19} & \msdd{1.58}{1.12} & \msdd{3.38}{1.74} & \msdd{5.28}{1.43} \\
ESP & \msdd{2.10}{1.20} & \msdd{5.96}{1.07} & \msdd{2.46}{1.48} & \msdd{2.34}{1.45} & \msdd{3.89}{1.71} & \msdd{5.37}{1.33} \\
DEU & \msdd{2.23}{1.26} & \msdd{5.72}{1.21} & \msdd{1.82}{1.31} & \msdd{1.83}{1.15} & \msdd{3.4}{1.67} & \msdd{5.15}{1.35} \\
BRA & \msdd{1.89}{0.97} & \msdd{6.28}{0.87} & \msdd{2.59}{1.60} & \msdd{2.38}{1.48} & \msdd{4.62}{1.71} & \msdd{5.98}{1.10} \\
RUS & \msdd{1.95}{1.10} & \msdd{5.93}{1.15} & \msdd{2.67}{1.67} & \msdd{2.83}{1.68} & \msdd{4.48}{1.61} & \msdd{5.60}{1.28} \\
MEX & \msdd{1.75}{1.09} & \msdd{6.18}{0.95} & \msdd{2.76}{1.53} & \msdd{2.51}{1.52} & \msdd{4.51}{1.65} & \msdd{5.92}{1.10} \\
IND & \msdd{1.71}{1.02} & \msdd{6.17}{1.02} & \msdd{2.54}{1.67} & \msdd{2.33}{1.49} & \msdd{4.56}{1.62} & \msdd{5.92}{1.12} \\
IDN & \msdd{1.99}{1.16} & \msdd{6.12}{0.88} & \msdd{3.06}{1.70} & \msdd{2.65}{1.43} & \msdd{4.62}{1.4} & \msdd{5.86}{1.11} \\
PHL & \msdd{1.53}{0.84} & \msdd{6.09}{1.05} & \msdd{2.71}{1.63} & \msdd{2.00}{1.30} & \msdd{4.63}{1.63} & \msdd{5.92}{1.09} \\
VEN & \msdd{1.31}{0.62} & \msdd{6.27}{0.98} & \msdd{2.76}{1.56} & \msdd{2.14}{1.43} & \msdd{4.97}{1.66} & \msdd{5.97}{1.09} \\

\bottomrule
\end{tabular}
%}
\label{table:means_sds}
\end{table}

\begin{table}[b!]
\caption{\textbf{Manifest correlations between factors.} This table shows the Pearson correlations between the observed scores of the six factors of the MCMS, calculated on the total sample (N = 5857). \mbox{$^{***}p < 0.001$}.
}

\centering

\sisetup{
            detect-all,
            table-number-alignment = center,
            table-figures-integer = 1,
            table-figures-decimal = 2,
            table-space-text-post = {\superscript{*}},
}

%\resizebox{\columnwidth}{!}{
\begin{tabular}{l S S S S S} 
\toprule
  & \textbf{Amotivation} & \textbf{Material} & \textbf{Social} & \textbf{Introjected} & \textbf{Identified}\\ 
\midrule
\textbf{Material}  & -0.21\textnormal{\superscript{***}} &  &  &  &  \\
\textbf{Social} & 0.02  & 0.12\textnormal{\superscript{***}} &  & &   \\
 \textbf{Introjected} & 0.13\textnormal{\superscript{***}} & 0.08\textnormal{\superscript{***}} & 0.52\textnormal{\superscript{***}} &  &   \\
\textbf{Identified} & -0.18\textnormal{\superscript{***}} & 0.35\textnormal{\superscript{***}} & 0.43\textnormal{\superscript{***}} & 0.39\textnormal{\superscript{***}} &  \\
\textbf{Intrinsic} & -0.43\textnormal{\superscript{***}} & 0.31\textnormal{\superscript{***}} & 0.28\textnormal{\superscript{***}} & 0.20\textnormal{\superscript{***}} & 0.46\textnormal{\superscript{***}}  \\
\bottomrule
\end{tabular}
%}
\label{table:factor_correlations}
\end{table}

We used Cronbach's alpha statistic \cite{cronbach1951coefficient} to assess the internal consistency of the MCMS. Table~\ref{table:cronbachs_alpha} displays the values of alpha for each country and income group. Alpha provides an estimate of the lower bound of internal consistency. As a rule of thumb, values above 0.7 are considered acceptable, values between 0.6 and 0.7 questionable, values between 0.5 and 0.6 poor and values below 0.5 unacceptable \cite{george2003using}. In most countries and groups, alpha exceeded 0.7 for each construct. Exceptions to this were the amotivation factor in Brazil and Venezuela as well as the material external regulation factor in Brazil and Indonesia with values between 0.5 and 0.7. Therefore, when interpreting results from these factors and groups, care should be taken.
In addition to Crohnbach's alpha, we calculated McDonald's coefficient omega \cite{mcdonald1999test} for assessing the reliability of the MCMS. The values for coefficient omega are shown in Appendix~\ref{app:omega}. Compared to Crohnbach's alpha, the values of coefficient omega are equal or slightly higher.

\begin{table}[b!]
\caption{\textbf{Internal consistency of the MCMS.} This table shows Cronbach's alpha values for all groups and constructs, along with a 95\% confidence interval for the values.}
\centering
%\resizebox{\textwidth}{!}{
\begin{tabular}{l c c c c c c} 
\toprule
  \textbf{Group} & \textbf{Amotivation} & \textbf{Material} & \textbf{Social} & \textbf{Introjected} & \textbf{Identified} & \textbf{Intrinsic}\\
\midrule
ALL & \ci{0.77}{0.78}{0.79} & \ci{0.77}{0.78}{0.79} & \ci{0.83}{0.84}{0.84} & \ci{0.82}{0.83}{0.84} & \ci{0.86}{0.87}{0.88} & \ci{0.88}{0.88}{0.89}\\ \midrule
HIGH & \ci{0.82}{0.84}{0.85} & \ci{0.80}{0.82}{0.83} & \ci{0.84}{0.85}{0.86} & \ci{0.85}{0.86}{0.87} & \ci{0.86}{0.87}{0.88} & \ci{0.89}{0.90}{0.91}\\
MID & \ci{0.68}{0.70}{0.73} & \ci{0.73}{0.75}{0.77} & \ci{0.81}{0.82}{0.84} & \ci{0.82}{0.83}{0.84} & \ci{0.85}{0.86}{0.87} & \ci{0.86}{0.87}{0.88}\\
LOW & \ci{0.76}{0.78}{0.80} & \ci{0.75}{0.77}{0.79} & \ci{0.84}{0.85}{0.87} & \ci{0.78}{0.80}{0.82} & \ci{0.83}{0.85}{0.86} & \ci{0.86}{0.87}{0.88}\\ \midrule
USA & \ci{0.84}{0.86}{0.87} & \ci{0.82}{0.84}{0.86} & \ci{0.81}{0.83}{0.85} & \ci{0.81}{0.83}{0.85} & \ci{0.84}{0.85}{0.87} & \ci{0.89}{0.90}{0.91}\\
ESP & \ci{0.81}{0.83}{0.86} & \ci{0.78}{0.80}{0.83} & \ci{0.83}{0.85}{0.87} & \ci{0.86}{0.87}{0.89} & \ci{0.88}{0.89}{0.91} & \ci{0.89}{0.90}{0.92}\\
DEU & \ci{0.80}{0.82}{0.85} & \ci{0.78}{0.80}{0.83} & \ci{0.86}{0.87}{0.89} & \ci{0.79}{0.82}{0.85} & \ci{0.82}{0.85}{0.87} & \ci{0.84}{0.86}{0.88}\\
BRA & \ci{0.45}{0.52}{0.59} & \ci{0.62}{0.67}{0.72} & \ci{0.8}{0.83}{0.85} & \ci{0.78}{0.81}{0.84} & \ci{0.87}{0.89}{0.90} & \ci{0.84}{0.86}{0.88}\\
RUS & \ci{0.74}{0.77}{0.80} & \ci{0.78}{0.81}{0.83} & \ci{0.87}{0.88}{0.90} & \ci{0.86}{0.88}{0.90} & \ci{0.83}{0.85}{0.87} & \ci{0.88}{0.89}{0.90}\\
MEX & \ci{0.76}{0.79}{0.82} & \ci{0.68}{0.72}{0.75} & \ci{0.74}{0.77}{0.80} & \ci{0.75}{0.78}{0.81} & \ci{0.83}{0.85}{0.87} & \ci{0.84}{0.86}{0.88}\\
IND & \ci{0.70}{0.74}{0.78} & \ci{0.79}{0.82}{0.84} & \ci{0.85}{0.87}{0.88} & \ci{0.76}{0.79}{0.82} & \ci{0.83}{0.85}{0.87} & \ci{0.86}{0.87}{0.89}\\
IDN & \ci{0.75}{0.79}{0.82} & \ci{0.59}{0.65}{0.71} & \ci{0.83}{0.85}{0.88} & \ci{0.75}{0.79}{0.82} & \ci{0.75}{0.79}{0.82} & \ci{0.82}{0.84}{0.87}\\
PHL & \ci{0.77}{0.80}{0.83} & \ci{0.77}{0.80}{0.83} & \ci{0.81}{0.84}{0.86} & \ci{0.79}{0.82}{0.84} & \ci{0.86}{0.88}{0.90} & \ci{0.87}{0.88}{0.90}\\
VEN & \ci{0.55}{0.60}{0.66} & \ci{0.74}{0.77}{0.80} & \ci{0.73}{0.76}{0.80} & \ci{0.73}{0.77}{0.80} & \ci{0.84}{0.86}{0.88} & \ci{0.81}{0.83}{0.86}\\
 \bottomrule
\end{tabular}
%}
\label{table:cronbachs_alpha}
\end{table}

\begin{table}[b!]
\caption{\textbf{Confirmatory Factor Analysis of the MCMS.} This table shows the goodness-of-fit statistics for the hypothesized MCMS model. The fit statistics are given for the total sample as well as for all groups.}
\centering
%\resizebox{\textwidth}{!}{
\begin{tabular}{lcccccccc} 
\hline
Group       & \textbf{N}  & \textbf{S-B}$\bm{\chi^2}$ & \textbf{df} & \textbf{CFI} & \textbf{TLI} & \textbf{RMSEA} & \textbf{RMSEA 90\% CI}   & \textbf{SRMR}  \\
\toprule
ALL & 5857 & 1590.49 & 120    & 0.965 & 0.955 & 0.046 & 0.044 \; 0.048 & 0.037 \\ \midrule
HIGH & 1952 & 573.64  & 120    & 0.970  & 0.961 & 0.044 & 0.041 \; 0.047 & 0.037 \\
MID & 1834 & 557.07  & 120    & 0.964 & 0.955 & 0.045 & 0.041 \; 0.048 & 0.038 \\
LOW & 1508 & 554.52  & 120    & 0.955 & 0.942 & 0.049 & 0.045 \; 0.053 & 0.036 \\\midrule
USA & 721  & 281.45  & 120    & 0.965 & 0.956 & 0.043 & 0.037 \; 0.049 & 0.040  \\
ESP & 677  & 272.66  & 120    & 0.975 & 0.968 & 0.043 & 0.037 \; 0.050 & 0.040  \\
DEU & 554  & 284.42  & 120    & 0.955 & 0.943 & 0.050  & 0.043 \; 0.056 & 0.044 \\
BRA & 496  & 256.58  & 120    & 0.957 & 0.946 & 0.048 & 0.040 \; 0.055 & 0.044 \\
RUS & 677  & 311.67  & 120    & 0.966 & 0.957 & 0.049 & 0.043 \; 0.055 & 0.039 \\
MEX & 661  & 316.27  & 120    & 0.948 & 0.933 & 0.050  & 0.043 \; 0.056 & 0.048 \\
IND & 608  & 272.66  & 120    & 0.962 & 0.951 & 0.046 & 0.039 \; 0.052 & 0.042 \\
IDN & 401  & 272.08  & 120    & 0.931 & 0.912 & 0.056 & 0.049 \; 0.064 & 0.050  \\
PHL & 499  & 291.47  & 120    & 0.954 & 0.941 & 0.054 & 0.046 \; 0.061 & 0.045 \\
VEN & 563  & 217.07  & 120    & 0.966 & 0.956 & 0.038 & 0.031 \; 0.045 & 0.039 \\ \bottomrule
\end{tabular}
%}
\label{table:cfa_fit}
\end{table}

\textbf{Confirmatory Factor Analysis.} 
In order to validate the factor structure of our hypothesized model, we conducted  a confirmatory factor analysis.
Table \ref{table:cfa_fit} shows the results of the confirmatory factor analysis of the hypothesized model. 
The first item of each factor in Table \ref{table:mcms_items} served as the marker variable (i.e., the loading of this item was fixed to one).
The analysis was conducted on the entire dataset as well as on each group separately. As in Section \ref{sec:existing}, we followed current conventions for evaluating model fit (e.g. \cite{kline2015principles, marsh2005goodness, hooper2008structural}).

The results show that the hypothesized model had adequate fit overall as well as in all groups. The CFI was above 0.95 in all groups except Indonesia and Mexico (with 0.931 and 0.948, respectively). RMSEA was lower than 0.06 in all groups and SRMR was lower than or equal to 0.05 in all groups. We consider the fit measures for Mexico and Indonesia to be marginally acceptable, but some care should be taken when interpreting the results from these countries.
Table \ref{table:model_parameters} shows the item loadings and intercepts estimated by the hypothesized model fitted to the entire sample and Table \ref{table:model_correlations} shows the estimated correlations between the constructs.
All estimated factor correlations are statistically significant at $p < 0.01$ (most at $p < 0.001$).

The alternative model which restricts the correlations of the external regulation factors (material external regulation and social external regulation) with intrinsic motivation to zero did not have an acceptable fit: While CFI was close to 0.95 for most groups (0.948 for the total sample), SRMR was high for the total sample (0.094) as well as for all other groups, ranging between 0.074 in the USA and 0.113 in Russia.

\begin{table}[b!]
\caption{\textbf{Estimated loadings and intercepts.} This table shows the standardized item loadings ($\lambda$) and the item intercepts ($\tau$) estimated by the hypothesized model (N = 5857). The first item of each construct is the marker item, with its (unstandardized) loading fixed at 1. The order of the items follows the order in Table~\ref{table:mcms_items}.}
\centering
%\resizebox{\textwidth}{!}{
\begin{tabular}{l c c c c c c c c c c c c} 
\toprule
   & \multicolumn{2}{c}{Amotivation} & \multicolumn{2}{c}{Material} & \multicolumn{2}{c}{Social} & \multicolumn{2}{c}{Introjected} & \multicolumn{2}{c}{Identified} & \multicolumn{2}{c}{Intrinsic} \\
   \hline  
      &  $\lambda$ & $\tau$ &   $\lambda$ & $\tau$ &  $\lambda$ & $\tau$ &  $\lambda$ & $\tau$ &  $\lambda$ & $\tau$ & $\lambda$ & $\tau$ \\   
\midrule 
item 1 & 0.775 & 1.824 & 0.823 & 5.987 & 0.894 & 2.285 & 0.787 & 2.251 & 0.829 & 4.274 & 0.868 & 5.623 \\
item 2 & 0.745 & 1.683 & 0.627 & 6.101 & 0.832 & 2.041 & 0.75  & 2.309 & 0.828 & 3.967 & 0.817 & 5.758 \\
item 3 & 0.700   & 2.012 & 0.776 & 6.059 & 0.71  & 3.086 & 0.826 & 2.186 & 0.845 & 4.56  & 0.855 & 5.637 \\
\bottomrule
\end{tabular}
%}
\label{table:model_parameters}
\end{table}

\begin{table}[b!]
\caption{\textbf{Variance extracted and shared variance.} This table shows the average variance extracted (AVE) and maximum shared variance (MSV) between the constructs. An AVE $\geq{0.5}$ indicates good convergence, and an AVE $>$ MSV is evidence for discriminant validity.}
\centering

\sisetup{
            detect-all,
            table-number-alignment = center,
            table-figures-integer = 1,
            table-figures-decimal = 3,
}

\begin{tabular}{l S S S S S S} 
\toprule

  & \textbf{Amotivation} & \textbf{Material} & \textbf{Social} & \textbf{Introjected} & \textbf{Identified} & \textbf{Intrinsic}\\ 
\midrule

AVE & 0.548 & 0.557 & 0.665 & 0.621 & 0.696 & 0.718 \\
MSV & 0.274 & 0.194 & 0.360 & 0.360 & 0.276 & 0.276\\

\bottomrule
\end{tabular}
\label{table:ave_msv}
\end{table}
\begin{table}[b!]
\caption{\textbf{Estimated correlations between constructs.} This table shows the model estimates of the correlations between constructs (N = 5857). $^{**}p < 0.01$, $^{***}p < 0.001$.}
\centering

\sisetup{
            detect-all,
            table-number-alignment = center,
            table-figures-integer = 1,
            table-figures-decimal = 3,
            table-space-text-post = {\superscript{*}},
}

%\resizebox{\columnwidth}{!}{
\begin{tabular}{l S S S S S} 
\toprule
  & \textbf{Amotivation} & \textbf{Material} & \textbf{Social} & \textbf{Introjected} & \textbf{Identified}\\ 
\midrule
\textbf{Material}  & -0.263\textnormal{\superscript{***}} &  &  &  &  \\

\textbf{Social} & 0.051\textnormal{\superscript{**}}  & 0.128\textnormal{\superscript{***}} &  & &   \\

 \textbf{Introjected} & 0.151\textnormal{\superscript{***}} & 0.108\textnormal{\superscript{***}} & 0.600\textnormal{\superscript{***}} &  &   \\
 
\textbf{Identified} & -0.222\textnormal{\superscript{***}} & 0.440\textnormal{\superscript{***}} & 0.458\textnormal{\superscript{***}} & 0.449\textnormal{\superscript{***}} &  \\

\textbf{Intrinsic} & -0.523\textnormal{\superscript{***}} & 0.362\textnormal{\superscript{***}} & 0.283\textnormal{\superscript{***}} & 0.230\textnormal{\superscript{***}} & 0.525\textnormal{\superscript{***}}  \\
\bottomrule
\end{tabular}
%}
\label{table:model_correlations}
\end{table}

Compared to other SDT-based approaches to measuring crowdworker motivations \cite{naderi2014development, naderi2018measure}, the MCMS achieved a better fit while measuring additional constructs. Other SDT-based scales used to measure motivations in the micro-task context \cite{naderi2014development, naderi2018measure} did not report a well-fitting model for sample sizes larger than 100 due to a high RMSEA ($> 0.08$ for the 12-item-subset of WEIMS \cite{naderi2014development} and $> 0.09$ for the Crowdwork Motivation Scale \cite{naderi2018measure}). For the Crowdwork Motivation Scale, the fit measures were better on two smaller samples (N = 90 and N = 86), but RMSEA was still high ($>0.072$ for both samples) and CFI was below 0.95.\footnote{TLI and SRMR were not reported.} In contrast, the MCMS achieved good fit measures on the total sample and for the individual countries (with CFIs above 0.95 for most countries and an RMSEA below 0.06 for all countries).

Compared to \texttt{A-WEIMS-M3}, the model that had the best fit in our evaluation of traditional work motivation scales, the MCMS showed a similar fit and suffers from none of \texttt{A-WEIMS-M3}'s drawbacks described in Section \ref{sec:existing}. Moreover, if only the four constructs measured by \texttt{A-WEIMS-M3} are taken into account, the MCMS achieves a better fit: A four-factor model of the MCMS with the intrinsic motivation and social external regulation factors ommited had a CFI of 0.983, a TLI of 0.984, an RMSEA of 0.037 and an SRMR of 0.026 on the total sample (N = 5857) and a CFI of 0.979, a TLI of 0.971, an RMSEA of 0.039 and an SRMR of 0.035 on the USA sample (N = 721).

\textbf{Construct Validity.} Construct validity refers to the extent to which the items of a scale ``accurately reflect the theoretical latent constructs they are designed to measure'' \cite{hair2018multivariate}. In the context of confirmatory factor analysis, a poor model fit is considered evidence of a lack of construct validity.
Besides achieving a good model fit, there are four additional components of construct validity that can be evaluated: content validity, convergent validity, discriminant validity and nomological validity \cite{hair2018multivariate}. Content validity was ensured during item development (see Section \ref{sec:development}). Evidence of the validity of the MCMS with respect to convergent, discriminant and nomological validity is discussed below. We follow the definitions of Hair et al. \cite{hair2018multivariate} for these types of validity.

\emph{Convergent Validity} is established if the items measuring a construct ``converge or share a high proportion of variance in common.'' \cite{hair2018multivariate} All factor loadings should be statistically significant and, especially for large samples, should be at least $\geq{0.5}$ (ideally, $\geq{0.7}$) \cite{hair2018multivariate}. Table \ref{table:model_parameters} shows the standardized factor loadings for each construct in the MCMS. All factor loadings, except for the item ExMat2 with a loading of 0.627, were $0.7$ or higher and all loadings were statistically significant at $p<0.001$, which provides evidence of convergent validity.

In addition to inspecting the factor loadings, the average variance extracted (AVE) can be used as a summary indicator of convergence. The AVE measures the average percentage of variation explained and is calculated as the mean of the squared standardized factor loadings of a construct. An AVE of $\geq{0.5}$ is evidence of good convergence \cite{hair2018multivariate, fornell1981evaluating}. Table \ref{table:ave_msv} shows the AVE for each construct. For all constructs, the AVE was $\geq{0.5}$, providing further evidence of convergent validity. 

\emph{Discriminant Validity} is ``the extent to which a construct or variable is truly distinct from other constructs or variables'' \cite{hair2018multivariate}. To establish discriminant validity, each construct's AVE should be greater than the squared correlation with other constructs, indicating that more variance in the construct's items is explained by the construct than the construct shares with other constructs  \cite{hair2018multivariate, fornell1981evaluating}. The highest squared correlation of a construct with any other construct, or maximum shared variance (MSV),  is shown in Table \ref{table:ave_msv}. 
For all constructs, the AVE was larger than the MSV, providing evidence of discriminant validity. 

\emph{Nomological Validity} is concerned with whether the relationships between the constructs of the scale correspond to theory and prior research \cite{hair2018multivariate}. Table \ref{table:model_correlations} shows the estimated correlations between the constructs of the MCMS.
As hypothesized by SDT and found in previous studies (e.g., \cite{weims, mwms}), we observe a negative correlation between intrinsic motivation and amotivation (as well as moderate negative correlation between identified regulation and amotivation). Furthermore, as hypothesized by SDT, the strongest correlations are between adjacent constructs (intrinsic motivation with identified regulation and introjected regulation with social external regulation), and generally\, constructs tend to correlate more strongly with adjacent constructs than with non-adjacent constructs.
There are exceptions to this pattern; notably concerning the material external regulation construct, which has a stronger correlation with intrinsic motivation and identified regulation than with adjacent constructs. However, some extent of deviation from the pattern of ordered correlations is not uncommon in SDT-based motivation scales (e.g. \cite{chemolli2014evidence, litalien2015motivation, howard2016using, noels2000you}) and research suggests that the hypothesized continuum is not necessarily represented well by the pattern of correlations estimated by CFA (e.g. \cite{howard2016using, litalien2015motivation}). Therefore, further research is necessary to determine the extent to which the MCMS follows the continuum structure hypothesized by SDT. 
Furthermore, while SDT hypothesizes that external regulation and intrinsic motivation are unrelated, previous research found a highly significant positive correlation between these constructs for both the MWMS (0.11) and the Academic Motivation Scale (0.49) \cite{chemolli2014evidence}. In line with these previous empirical results, our results also show a positive correlation between intrinsic motivation and both material and social external regulation.

\textbf{Criterion Validity.} For a first analysis of the relationship between motivational dimensions and the behavioral outcomes of crowdworkers, we investigated the relationship between the scores of the motivational constructs and (1) the time taken to complete the task, (2) the amount of text content that crowdworkers produced in response to the open-ended question, and (3) the self-reported time spent on CrowdFlower per week.
We use the first two measures (\emph{time taken} and \emph{text produced}) as an estimate for the effort that workers put into the task. 

Previous empirical research has found that amotivation correlates negatively with effort \cite{mwms}, that autonomous motivation\footnote{Autonomous motivation encompasses identified regulation and intrinsic motivation (e.g., \cite{mwms, sheldon1998not}).}  correlates more positively with effort than controlled motivation\footnote{Controlled motivation encompasses external regulation and introjected regulation (e.g., \cite{mwms, sheldon1998not}).} \cite{mwms, sheldon1998not} and that autonomous motivation predicts greater effort \cite{sheldon1998not}. These previous findings are consistent with SDT: For intrinsic motivation, the positive emotions associated with enjoying an activity naturally reinforce persistent effort, and identified regulation may elicit such reinforcing emotions due to value congruence even if the activity itself is not enjoyable \cite{sheldon1998not}. In line with previous research \cite{mwms} and consistent with SDT, we expected effort to be correlated negatively with amotivation and positively with autonomous motivation. Furthermore, we expected the correlation of effort with autonomous motivation to be more positive than the correlation of effort with controlled motivation.

Concerning the weekly time spent on the platform (\emph{weekly time}), in line with previous research on crowdworker motivations \cite{kaufmann2011more}, we expected the motivational profiles to differ between workers who spend a lot of time on the platform and workers who spend little time on the platform. Specifically, Kaufmann et al. \cite{kaufmann2011more} found that skill variety\footnote{Kaufmann et al. \cite{kaufmann2011more} defined skill variety as ``usage of a diversity of skills that are needed for
solving a specific task and fit with the skill set
of the worker,'' e.g. a worker picking ``a translation
task because he likes translating.'' In the SDT context, this construct contains aspects of intrinsic motivation and the fulfillment of the need for competence.}, human capital advancement\footnote{Kaufmann et al. \cite{kaufmann2011more} defined human capital advancement as ``motivation through the possibility
to train skills that could be useful to
generate future material advantages,'' e.g. a worker choosing a task ``because he or she wants
to improve language skills for a new or better job,'' which can be interpreted as an aspect of identified regulation.} and community identification\footnote{Kaufmann et al. \cite{kaufmann2011more} defined community identification as a ``subconscious adoption of norms and values from the crowdsourcing platform community, which is caused by a personal identification
process.'' In the SDT context, this construct is most similar to a regulation that has been internalized to a great extent, i.e. identified or integrated regulation.} correlated positively with weekly time spent, while the correlation with pastime\footnote{Kaufmann et al. \cite{kaufmann2011more} defined pastime as ``acting just to `kill time,'' e.g. a worker who ``works on various `random' tasks because he has nothing
better to do.'' In the context of SDT, this could be interpreted as amotivation.} was negative. Furthermore, in the context of SDT, previous research (e.g., \cite{deci2002handbook}) found that intrinsic motivation and highly internalized extrinsic motivation was associated with longer persistence in an activity.
Therefore, we expected amotivation to be negatively correlated with weekly time spent, and we expected the strongest positive correlations to be with identified regulation and intrinsic motivation.

\begin{table}[b!]
\caption{\textbf{Correlations between motivational constructs and estimates of effort:} This figure shows the Pearson correlations between the different types of motivations and two estimates or effort: the time taken to complete the task and the amount of text content produced by the workers. Furthermore, it shows the Spearman correlations between the different types of motivation and weekly time spent on the platform. $^{*}p < 0.05$, $^{**}p < 0.01$, $^{***}p < 0.001$.}
\centering

\sisetup{
            detect-all,
            table-number-alignment = center,
            table-figures-integer = 1,
            table-figures-decimal = 2,
            table-space-text-post = {\superscript{*}},
}

\begin{tabular}{l S S S S S S} 
\toprule

  & \textbf{Amotivation} & \textbf{Material} & \textbf{Social} & \textbf{Introjected} & \textbf{Identified} & \textbf{Intrinsic}\\ 
\midrule

time taken & -0.14\textnormal{\superscript{***}} & 0.04\textnormal{\superscript{**}} & 0.09\textnormal{\superscript{***}} & 0.10\textnormal{\superscript{***}} & 0.18\textnormal{\superscript{***}} & 0.15\textnormal{\superscript{***}} \\
text produced  & -0.14\textnormal{\superscript{***}} & 0.06\textnormal{\superscript{***}} & 0.00\textnormal{\superscript{}} & 0.03\textnormal{\superscript{*}} & 0.11\textnormal{\superscript{***}} & 0.10\textnormal{\superscript{***}} \\
weekly time  & -0.15\textnormal{\superscript{***}} & 0.13\textnormal{\superscript{***}} & 0.14\textnormal{\superscript{***}} & 0.08\textnormal{\superscript{***}} & 0.21\textnormal{\superscript{***}} & 0.24\textnormal{\superscript{***}} \\

\bottomrule
\end{tabular}
\label{table:time_character}
\end{table}

To calculate the time taken to complete the task, we used the starting and finishing times reported by the platform, and for an estimate of amount of text content that a worker had produced, we counted the characters typed after removal of stopwords. 
Table~\ref{table:time_character} shows the Pearson correlations between the different types of motivation and the two estimates of effort.\footnote{We logarithmized the character count as the distribution was heavily skewed.} While our estimates are noisy operationalizations of job effort (e.g. due to differences in reading and typing skills), they measure observed behavior as opposed to self-reports. 

Our results support the hypothesis regarding effort and we observe the same pattern as found in previous research \cite{mwms} via self-reported job effort: Both measures have a significant negative correlation with amotivation and significant positive correlations with autonomous types of motivation (identified regulation and intrinsic motivation). Furthermore, the time taken correlates positively with controlled motivation types, and the amount of text produced correlates positively with material external regulation and introjected regulation. As hypothesized, the correlations between effort and the autonomous motivation are more positive than the correlations between effort and controlled motivation. All differences in correlation strength between the types of autonomous motivation and the types of controlled motivation are statistically significant.\footnote{To assess the statistical significance of the differences, we used Steiger’s method \cite{steiger1980tests} for statistical comparisons between correlations measured on the same sample, as implemented by Lee and Preacher \cite{lee2013calculation}. Our results showed that the correlations between the types of autonomous motivation and both measures of task effort were significantly stronger than the correlations between these measures and the types of controlled motivation. The differences were significant at p < 0.01 or lower.} 

To gather data on the weekly time spent on CrowdFlower, we asked workers the question \emph{``How much time do you spend on CrowdFlower, per week?''} and workers were given seven answer possibilities, ranging from ``less than 1 hour'' to ``more than 40 hours.'' This question was included in the demographics part of the task. Table~\ref{table:time_character} shows the correlations between the motivational constructs and the weekly time spent on CrowdFlower. Due to the ordinal nature of this variable, we report Spearman correlations. In line with previous research on SDT as well as previous research on crowdworker motivations, we find the highest positive correlations of weekly time spent with identified regulation and intrinsic motivation, and we find a negative correlation with amotivation.

\textbf{Applicability Across Platforms.} For a first evaluation of the extent to which the MCMS validly measures motivations of crowdworkers on other micro-task platforms, we administered the MCMS to a small sample of crowdworkers on AMT. For use on AMT, we substituted the term ``CrowdFlower tasks'' with ``tasks on Amazon Mechanical Turk'' in the stem and items of the MCMS. We collected 150 responses from Indian workers on AMT in June 2017. After spam removal, the sample contained 109 responses. CFA results on this sample showed good fit ($CFI = 0.961$, $TLI = 0.951$, $RMSEA = 0.049$, $SRMR = 0.068$). Furthermore, measurement invariance tests (see Section~\ref{sec:invariance} for more details on measurement invariance) showed that the AMT sample had scalar invariance with the sample of Indian crowdworkers on CrowdFlower. These results indicate that not only is the MCMS likely to be valid on other platforms, but that the MCMS also likely allows for a valid comparison of group means across the platforms.

In sum, our results indicate that the MCMS is a reliable and valid measurement of crowdworker motivations within the framework of self-determination theory. Researchers wishing to measure the motivation of crowdworkers, for example along with other variables such as behavioral patterns of crowdworkers, can easily include the scale as a module in their task design. Instruction for use of the MCMS are given in Appendix~\ref{app:scale_use}.

\section{Cross-Group Comparability of MCMS Results}
\label{sec:invariance}

When comparing the results of a measurement instrument across different groups, it is important to ensure that the instrument possesses the same psychometric properties in all groups. This characteristic is referred to as \emph{measurement invariance}. Tests of measurement invariance evaluate ``whether or not, under different conditions of observing and studying phenomena, measurement operations yield measures of the same attribute'' \cite{horn1992practical}. In our case, measurement invariance means that crowdworkers from different countries (or country income groups) assign the same meaning to the items used in the MCMS. Measurement invariance is particularly important if mean level differences between countries ought to be compared. A lack of measurement invariance can indicate, for example, that respondents of different groups understand the items in a different way (e.g. due to culture) or that different levels of response biases are present (e.g. \cite{cheung2000assessing}).

To evaluate measurement invariance of the MCMS, we conducted multiple-group confirmatory factor analyses (MGCFA). In MGCFA, measurement invariance is evaluated via a series of hypothesis tests which test invariance at different levels. 
Three levels of measurement invariance are commonly tested: configural, metric and scalar invariance (e.g. \cite{cheung2002evaluating, chen2007sensitivity}). 
Configural invariance requires that the items share the same configurations of loadings in all groups. 
Metric invariance additionally requires that the loadings of each item on its factor is the same across groups. 
Scalar invariance additionally requires that the intercepts of item regressions on each factor are the same across groups. To validly compare manifest mean differences across groups, scalar invariance is required. To validly compare latent mean differences across groups, at least partial scalar invariance is required \cite{millsap2012investigating}.

\begin{table}[b!]
\caption{\textbf{Measurement invariance.} This table shows the results of the MGCFA tests of invariance between groups and countries. The deltas are with respect to the previous level of measurement invariance, i.e. for scalar invariance, the sum of the deltas for metric and scalar invariance should be below 0.01 for CFI and below 0.015 for RMSEA. 
* 1 free intercept (\emph{Am3}), ** 5 free intercepts (\emph{Am3}, \emph{ExMat2}, \emph{ExSoc3}, \emph{Introj2} and \emph{Ident2}).}
\centering
%\scalebox{.8}{
\begin{tabular}{l l c c c} 
\toprule
 & \textbf{CFI} & \textbf{CFI} $\Delta$ & \textbf{RMSEA} & \textbf{RMSEA} $\Delta$ \\  
 \midrule
 
 \textbf{Income Groups} & & & & \\
    \tab Configural Invariance      & 0.964 & n/a    & 0.046 & n/a    \\
    \tab Metric Invariance          & 0.963 & 0.001 & 0.045 & 0.001 \\
    \tab Full Scalar Invariance     & 0.952 & 0.011 & 0.049 & 0.005 \\
    \tab Partial Scalar Invariance* & 0.955 & 0.008 & 0.048 & 0.004 \\
 \midrule

 \textbf{Countries} & & & & \\
    \tab Configural Invariance       & 0.960  & n/a    & 0.047 & n/a   \\
    \tab Metric Invariance           & 0.959 & 0.001 & 0.046 & 0.001 \\
    \tab Full Scalar Invariance      & 0.930  & 0.028 & 0.058 & 0.011 \\
    \tab Partial Scalar Invariance** & 0.952 & 0.007 & 0.049 & 0.002 \\
 \bottomrule
\end{tabular}
%}
\label{table:invariance}
\end{table}

We tested configural, metric and scalar invariance of the MCMS across countries and across income groups.
Configural invariance is indicated by acceptable goodness-of-fit indices in an MGCFA model without any equality constraints \cite{van2012checklist}.
For indications of metric and scalar non-invariance, we follow the guidelines of Chen \cite{chen2007sensitivity}:
For metric and scalar invariance tests on large samples ($N > 300$), a change of $\geq -0.010$ in CFI supplemented by a change of $\geq 0.015$ in RMSEA indicates non-invariance.

Table \ref{table:invariance} shows the goodness-of-fit indices for the progressively restricted models.
The results show a good fit for the model without equality constraints, indicating that configural invariance holds. 
Full metric invariance was also achieved, indicating that the strength of the relationship between the items and constructs is the same across groups.
Full scalar invariance could not be achieved. However, partial scalar invariance was achieved by releasing one intercept for the income groups (\emph{Am3}) and five intercepts in the countries (\emph{Am3}, \emph{ExMat2}, \emph{ExSoc3}, \emph{Introj2} and \emph{Ident2}).
Partial scalar invariance still allows for factor means to be compared as long as at least two intercepts per factor are invariant. Care should be taken, however, as Steinmetz' \cite{steinmetz2013analyzing} simulations showed that unequal intercepts may lead to erroneous conclusions about mean-level differences when comparing observed composite means across groups.
Therefore, a cross-country or cross-income group comparison of crowdworker motivations measured with the MCMS should rely on the model-implied latent means, which take intercept non-invariance into account, instead of observed composite means.

In sum, our analyses showed that the MCMS is well-suited for measuring the motivations of crowdworkers based in different countries. Furthermore, the invariance analyses indicated that the motivations measured with the MCMS can be used for a comparison of motivations across countries and across country income groups. 

\section{Motivations of Crowdworkers on CrowdFlower}
\label{sec:results}

This section reports the motivations of crowdworkers on CrowdFlower as measured by the MCMS as well as the results of our cross-country and cross-income group comparison of crowdworker motivations.

\textbf{Motivations Measured with the MCMS.} Table~\ref{table:latent_means} shows the latent means\footnote{To obtain estimates of the latent means, we fixed the marker items' intercepts to zero.} of the different motivational dimensions for the entire sample (\texttt{ALL}). 
Our results show that overall, material external regulation was the most important motivation for crowdworkers, with a mean of 5.99, followed by intrinsic motivation with a mean of 5.62. This points to an interesting duality of monetary and interest-driven, enjoyment-based motivational influences. The result is consistent with
previous research on crowdworker motivations on AMT (e.g., \cite{kaufmann2011more}), which found that both payment and enjoyment play an important role for crowdworkers, with monetary reasons being slightly more important than enjoyment.
The construct with the third highest mean was identified regulation (mean 4.27), which signifies that putting effort into CrowdFlower tasks is moderately in alignment with crowdworkers' personal goals and objectives such as lifestyle preferences or career plans. 
Social external regulation, introjected regulation and amotivation were the least important motivational factors for crowdworkers overall, with amotivation having the lowest score of all motivational dimensions (means 2.29, 2.25 and 1.82, respectively). 

As an illustration of how the different motivational types can be interpreted in the micro-task context, we give examples of the reasons that crowdworkers have for doing CrowdFlower tasks in the workers' own words. 
In the first section of the task, we instructed crowdworkers to think of five reasons for why they do tasks on CrowdFlower and asked them to write down these reasons. For each construct, Table~\ref{table:text-examples} shows examples of answers to this question that can be interpreted to correspond to the theoretical definition of the construct. The examples given are taken, for each construct, from four different crowdworkers who had a mean score greater than $5$ on the respective construct.

\begin{table}[b!]
\caption{\textbf{Responses to the open-ended question.} This table shows illustrative examples of answers to the open-ended question ``Give 5 reasons for why you do CrowdFlower tasks.'' from crowdworkers who had a mean score of $>5$ on the respective construct.}
\centering
\begin{tabular}{l p{10.5cm}} 
\toprule
\textbf{Construct} & \textbf{Examples}\\
\midrule
 \textbf{Amotivation} & \textit{``I'm bored'', ``I have nothing else to do'', ``this iss boring'', ``Nothing''}\\[1px]

 \textbf{Material} & \textit{``I need the money'', ``I get paid'', ``To get an extra income!'', ``good profit''}\\[1px]

 \textbf{Social} & \textit{``other people want me to fulfill the job'', ``my friends do it too'', ``referral from a friend'', ``friends do it''}\\[1px]

 \textbf{Introjected} & \textit{ ``I'm no worse than others'', ``To feel better about myself'', ``I feel useful'', ``They make me feel productive and useful.''}\\[1px]

 \textbf{Identified} & \textit{``To be my own boss!'', ``To gain some new experience'',  ``It helps me improve my English'', ``I can work from anywhere''}\\[1px]

 \textbf{Intrinsic} & \textit{``Enjoy seeing a variety of different topics'', ``Most of the tasks are actually entertaining'', ``There are tasks that are very interesting'', ``I enjoy sharing thoughts and opinions while completing CrowdFlower tasks.''}\\
    
 \bottomrule
\end{tabular}
\label{table:text-examples}
\end{table}

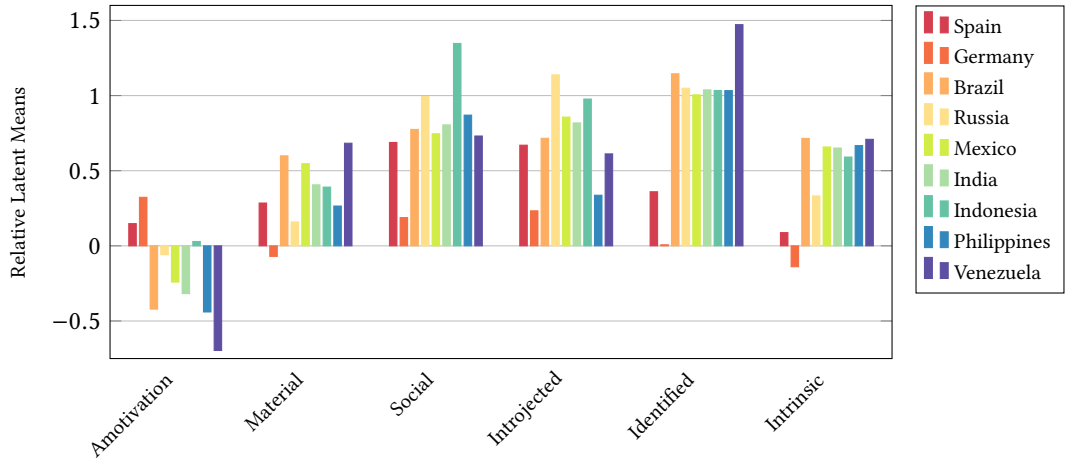
\begin{figure}[t]
	\centering
	\begin{center}
            \resizebox{0.9\textwidth}{!}{
\begin{tikzpicture}
    \begin{axis}[
        width  = 12cm,
        height = 6.29cm,
        major x tick style = transparent,
        ybar=1.2pt,
        bar width=0.1cm,
        ymax=1.6pt,
        ymin=-.75pt,
        ymajorgrids = true,
        ylabel = {Relative Latent Means},
        symbolic x coords={Amotivation,Material,Social,Introjected,Identified,Intrinsic},
        xtick = data,
        ytick = {-0.5,0.0,0.5,1.0,1.5},
       nodes near coords align={vertical},
       x tick label style={rotate=45,anchor=east,font=\footnotesize},
       ylabel style={font=\footnotesize},
       legend style={font=\footnotesize},
       scaled y ticks = false,
       legend pos=outer north east,
       legend cell align={left},
    ]

        \addplot[style={cspain,fill=cspain,mark=none}]
             coordinates {(Amotivation, 0.148) (Material,0.285) (Social, 0.688)(Introjected, 0.67) (Identified, 0.36)(Intrinsic, 0.088)};
             
        \addplot[style={cgermany,fill=cgermany,mark=none}]
             coordinates {(Amotivation, 0.323 ) (Material,-0.07) (Social, 0.188)(Introjected, 0.233) (Identified, 0.007)(Intrinsic, -0.139)};

        \addplot[style={cbrazil,fill=cbrazil,mark=none}]
             coordinates {(Amotivation,  -0.421) (Material,0.599) (Social, 0.775)(Introjected,0.716) (Identified, 1.145)(Intrinsic, 0.715)};

        \addplot[style={crussia,fill=crussia,mark=none}]
             coordinates {(Amotivation,  -0.059) (Material,0.159 ) (Social,0.994 )(Introjected,1.138 ) (Identified, 1.049)(Intrinsic, 0.332)};
             
        \addplot[style={cmexico,fill=cmexico,mark=none}]
             coordinates {(Amotivation, -0.241 ) (Material,0.547) (Social,0.746 )(Introjected, 0.857) (Identified,1.005  )(Intrinsic, 0.658)};

        \addplot[style={cindia,fill=cindia,mark=none}]
             coordinates {(Amotivation, -0.318 ) (Material,0.406) (Social, 0.805)(Introjected, 0.818) (Identified, 1.038)(Intrinsic, 0.651)};

        \addplot[style={cindonesia,fill=cindonesia,mark=none}]
             coordinates {(Amotivation, 0.029 ) (Material,0.391) (Social, 1.347)(Introjected, 0.977) (Identified,1.034 )(Intrinsic, 0.591)};

        \addplot[style={cphilipines,fill=cphilipines,mark=none}]
             coordinates {(Amotivation, -0.44 ) (Material,0.265) (Social, 0.87)(Introjected, 0.337) (Identified,1.033 )(Intrinsic, 0.667)};

        \addplot[style={cvenezuela,fill=cvenezuela,mark=none}]
             coordinates {(Amotivation,-0.696  ) (Material,0.683) (Social, 0.731)(Introjected, 0.612) (Identified,1.472)(Intrinsic,0.709)};

        \legend{Spain,Germany,Brazil,Russia,Mexico,India,Indonesia,Philippines,Venezuela}
    \end{axis}
\end{tikzpicture}
}
	\end{center}	
  	\caption{\textbf{Country differences in latent means, compared to the USA sample.} This figure shows the differences in latent means for all constructs and countries, compared to the latent means of the U.S. sample.}
  \centering
  \label{fig:differences}
\end{figure}

\textbf{Differences in Motivations across Groups.} 
As partial scalar invariance was achieved, the analysis of group differences in motivations measured with the MCMS relies on latent means estimated by the model instead of observed means. For analyzing the differences in latent constructs between groups, one group was chosen as the reference group. For our analysis, we chose the high income group sample as the reference group for the cross-income group comparison and the USA sample as the reference group for the cross-country comparison. Figure \ref{fig:differences} shows the differences in latent means of the different countries, compared to the reference group (USA), and Table~\ref{table:latent_means} shows the latent means for all groups, as well as the mean differences to the reference group in parentheses.\footnote{We obtained the estimated differences and their statistical significances via MGCFA with the means of the reference group fixed to zero. The threshold for all reported significances is set at $p < 0.001$, except amotivation in Spain, external social regulation in Germany and external material regulation in Russia ($p < 0.05$).}
 
Regarding the ranks of the constructs, the results show that the ranks of the three constructs with the highest scores were the same across all groups.
External material regulation was the construct with the highest score, followed by intrinsic motivation with only a slightly lower score, in all countries and income groups. Furthermore, the third most important motivational factor was identified regulation in all groups. The other motivational dimensions differed in rank across countries and income groups. Amotivation was the construct with the lowest score in the middle and low income groups, while in the high income group, it had a higher score than social external regulation and introjected regulation.
 
Regarding the differences in construct scores, the results show that motivations differ significantly between crowdworkers of different countries and country income groups.\footnote{Note that, as described in Section~\ref{sec:validation}, some care should be taken when interpreting results from Mexico and Indonesia due to a CFI lower than 0.95 (but above 0.90) and when interpreting the amotivation construct in Brazil and Venezuela, as well as the material external regulation construct in Brazil and Indonesia due to Crohnbach's alpha values below 0.7.} The largest differences in motivation, compared to workers in the USA, are with countries that are in income groups lower than the USA.
 
Amotivation was significantly higher in the high income group than in the middle and low income groups. Crowdworkers in Brazil, Mexico, India, the Philippines and Venezuela had a significantly lower amotivation score than U.S workers, while German and Spanish crowdworkers exhibited a significantly higher level of amotivation than workers in the USA. 
This indicates that in high income countries, crowdworkers tend to perceive doing CrowdFlower tasks as more ``pointless'' and a ``waste of time'' than in most lower income countries. 

\begin{table}[b!]
\caption{\textbf{Latent scale means and group differences.} This table shows the latent means for all groups, as well as the estimated differences in latent means in parentheses.}

\centering

\sisetup{
    table-align-text-pre     = false,
    table-align-text-post    = false,
    input-signs              = + -,
    input-open-uncertainty   = ,
    input-close-uncertainty  = 
}

\begin{tabular}{
l 
S[table-format=1.2]@{}
S
S[table-format=1.2]@{\;}
S
S[table-format=1.2]@{}
S
S[table-format=1.2]@{}
S
S[table-format=1.2]@{}
S
S[table-format=1.2]@{\;}
S
} 
\toprule
  \textbf{Group} & 
  \multicolumn{2}{l}{\textbf{Amotivation}} & \multicolumn{2}{l}{\textbf{Material}} & \multicolumn{2}{l}{\textbf{Social}} & \multicolumn{2}{l}{\textbf{Introjected}} & \multicolumn{2}{l}{\textbf{Identified}} & \multicolumn{2}{l}{\textbf{Intrinsic}}\\
\midrule
ALL & 1.82 && 5.99 && 2.29 && 2.25 && 4.27 &&  5.62 &\\ 
\midrule
HIGH & 2.12 & &  5.75 &&  1.86 &&  1.92 &&  3.61 &&  5.22 & \\
MID & \msddb{1.75}{-0.36} &  \msddb{6.08}{0.32} &  \msddb{2.50}{0.64} &  \msddb{2.57}{0.65} &  \msddb{4.54}{0.93} &  \msddb{5.78}{0.56} \\
LOW & \msddb{1.71}{-0.41} &  \msddb{6.06}{0.30} &  \msddb{2.58}{0.73} &  \msddb{2.30}{0.38} &  \msddb{4.6}{0.99} &  \msddb{5.87}{0.65} \\
\midrule
USA & 1.97 && 5.68 &&  1.60 && 1.62 && 3.49 && 5.23 & \\
ESP & \msddb{2.12}{0.15} &  \msddb{5.97}{0.28} &  \msddb{2.29}{0.69} &  \msddb{2.29}{0.67} &  \msddb{3.85}{0.36} &  \msddb{5.31}{0.09} \\
DEU & \msddb{2.30}{0.32} &  \msddb{5.61}{-0.07} &  \msddb{1.79}{0.19} &  \msddb{1.85}{0.23} &  \msddb{3.5}{0.01} &  \msddb{5.09}{-0.14} \\
BRA & \msddb{1.55}{-0.42} &  \msddb{6.28}{0.60} &  \msddb{2.38}{0.78} &  \msddb{2.34}{0.72} &  \msddb{4.64}{1.15} &  \msddb{5.94}{0.72} \\
RUS & \msddb{1.92}{-0.06} &  \msddb{5.84}{0.16} &  \msddb{2.60}{0.99} &  \msddb{2.76}{1.14} &  \msddb{4.54}{1.05} &  \msddb{5.56}{0.33} \\
MEX & \msddb{1.73}{-0.24} &  \msddb{6.23}{0.55} &  \msddb{2.35}{0.75} &  \msddb{2.48}{0.86} &  \msddb{4.50}{1.00} &  \msddb{5.88}{0.66} \\
IND & \msddb{1.66}{-0.32} &  \msddb{6.09}{0.41} &  \msddb{2.41}{0.80} &  \msddb{2.44}{0.82} &  \msddb{4.53}{1.04} &  \msddb{5.88}{0.65} \\
IDN & \msddb{2.00}{0.03} &  \msddb{6.07}{0.39} &  \msddb{2.95}{1.35} &  \msddb{2.60}{0.98} &  \msddb{4.53}{1.03} &  \msddb{5.82}{0.59} \\
PHL & \msddb{1.53}{-0.44} &  \msddb{5.95}{0.26} &  \msddb{2.47}{0.87} &  \msddb{1.96}{0.34} &  \msddb{4.53}{1.03} &  \msddb{5.89}{0.67} \\
VEN & \msddb{1.28}{-0.70} &  \msddb{6.36}{0.68} &  \msddb{2.33}{0.73} &  \msddb{2.23}{0.61} &  \msddb{4.96}{1.47} &  \msddb{5.93}{0.71} \\

\bottomrule
\end{tabular}

\label{table:latent_means}
\end{table}

While material external regulation had the highest score
of all constructs in all countries, scores were significantly higher in some countries than in others. Germany was the country with the lowest score on this construct, while Venezuela had the highest score. Both the middle and the low income group reported higher scores for material external regulation than the high income group, and crowdworkers of all countries except Germany reported a significantly higher material external regulation than U.S. workers. This means that workers in countries with lower incomes tend to be more motivated by the material rewards of micro tasks than workers in high income countries.

Social external regulation and introjected regulation scores were significantly higher in the middle and low income groups than in the high income group, and significantly higher in all countries, compared to the USA sample. This means that satisfying external social demands, as well as the avoidance of shame or guilt feelings, are more important motivational factors for workers in low and middle income countries and in countries other than the USA.

The largest differences in construct scores, both between countries and between income groups, were found for identified regulation. Scores were significantly higher in countries of the middle and low income groups than in countries of the high income group. This means that in lower income countries, crowdworkers perceive putting effort into micro tasks as more in line with their personal goals, objectives and values. One reason for this might be that in higher income countries, the same goals can be achieved more effectively through other means. Crowdworkers in the USA had the lowest identified regulation score of all countries and scores were significantly higher in all other countries except Germany.

Finally, all countries except Spain and Germany reported a significantly higher intrinsic motivation than workers in the USA, and middle and low income groups reported a higher intrinsic motivation score than the high income group. This means that that workers of countries in the middle and low income groups are more driven by interest and enjoyment inherent in the activity.

\section{Conclusion}
\label{sec:conclusion}

In this paper, we developed and validated the Multidimensional Crowdworker Motivation Scale (MCMS), a new scale for measuring crowdworker motivations. The scale measures the motivations of crowdworkers on six dimensions, based on the conceptualization of motivation suggested by self-determination theory. 
To the best of our knowledge, the MCMS is the first instrument developed specifically for measuring motivation in the micro-task context that provides a comprehensive representation of the motivational dimensions hypothesized by self-determination theory. Compared to existing instruments, the MCMS allows for a more comprehensive and theoretically well-founded measurement of crowdworker motivations with only three items per motivational dimension. Moreover, it is the first instrument for measuring crowdworker motivations that is validated in multiple countries and income groups.

The hypothesized six-factor MCMS model generally showed good fit in all countries and income groups. In addition, the results of the measurement invariance tests demonstrated that partial scalar invariance holds. This implies that the substantive meaning of the motivations measured with the MCMS are comparable across countries and income groups, allowing for valid cross-national comparisons. By providing a reliable scale for measuring crowdworker motivations, our study constitutes an important step towards a better understanding of the international crowd-workforce.

We designed the MCMS with generalizability across platforms in mind. By exchanging the platform name in the instructions and items of the MCMS, the scale is applicable for measuring the motivations of crowdworkers on other micro-task platforms. Moreover, it likely allows for valid cross-platform comparisons of motivations, as we demonstrated with a sample of workers on AMT.

Finally, in this paper we have presented a first cross-country and income group comparison of crowdworker motivations on the micro-task platform CrowdFlower.
This data provides novel insights regarding wide-ranging differences of the motivations for participating on such a platform. 

Given the good psychometric properties of the MCMS, future research on crowdwork can utilize the scale in order to address substantive questions concerning crowd employment. Here, the six motivational dimensions could serve as an outcome or as an explanatory variable. Potential fields of application\footnote{For instructions on use of the MCMS, see Appendix~\ref{app:scale_use}.} include predicting worker retention, investigating the relationship between worker motivation and productivity in different tasks, and conducting comparison studies of the motivations of different crowdworker populations, among others. Furthermore, the MCMS contributes to answering the question as to where in the employment space crowd employment should be located.
The MCMS is also relevant for micro-task platform developers who can use the scale to assess whether changes made to the platform lead to desirable or undesirable changes in motivation.

The work presented in this paper has several limitations. First of all, the scale was presented in English to the crowdworkers in all countries, which means that we are only able to capture the motivations of crowdworkers with appropriate English skills. However, we can assume that a majority of crowdworkers on CrowdFlower possess an adequate level of English skills, as the platform interface is available exclusively in English and workers are expected to understand instructions in English. Furthermore, demand for crowdworkers is driven by English-speaking countries \cite{kuek2015global}. 
Regarding the presence of social desirability bias, Blais et al. \cite{blais1993work} found that self-reported work motivations only correlated very weakly with the Marlow-Crowne Social Desirability Scale \cite{crowne1960new}. However, as Antin and Shaw \cite{antin2012social} found evidence for the presence of social desirability bias in self-reported motivations of crowdworkers, further experiments are needed in order to assess the extent to which social desirability bias is present in data collected with the MCMS.
A further limitation of the MCMS is that it does not measure integrated regulation. The lack of an integrated regulation factor in the MCMS is due to problems of statistically distinguishing this factor from identified regulation and intrinsic motivation.
Due to the same problems, this limitation also applies to other SDT-based work motivation scales such as the MWMS \cite{mwms}. Finally, the MCMS was developed specifically for the context of paid micro tasks and is not intended for use in other crowdsourcing contexts. While an application in other contexts may be possible, the scale would have to be adapted first (e.g. by removing the material external regulation construct for application in the context of unpaid tasks) and validated in the respective context.

In future work, we plan to conduct a more in-depth analysis of cross-country and cross-income group differences, including their stability over time. Additionally, we plan to further investigate the relationship between motivations and economic as well as demographic factors, going beyond the country of residence as an indicator of difference. 
Regarding cross-platform comparability of MCMS responses, we plan to further evaluate the MCMS on other micro-task platforms and, provided that measurement invariance is achieved, conduct a cross-platform analysis of worker motivations.

Another direction that we plan to follow in future work is an evaluation of the extent to which motivations measured with the MCMS are related to different antecedents and outcomes. For investigating the relations between measured motivations and antecedents (e.g. the satisfaction of basic needs) or outcomes (e.g. emotional exhaustion), the scales for measuring the antecedents and outcomes will first have to be validated within the crowdworking domain. 
Finally, in future work we plan to develop Bayesian models that incorporate not only the responses to the MCMS items but also responses to open-ended survey questions and demographic metadata. 

To conclude, the MCMS constitutes a promising step forward in measuring the motivations of crowdworkers in a theoretically founded, reliable and internationally comparable way. 
By shedding light on the motivations of the ``indefinite and unknown group'' of crowdworkers, the MCMS enables novel insights into this emerging form of short-term employment.
This work is relevant not only for researchers but also for practitioners seeking to measure the motivations of crowdworkers and to harness knowledge about differences in crowdworker motivations.

\bibliographystyle{ACM-Reference-Format}
\bibliography{references}

\clearpage

\appendix

\section{The Multidimensional Crowdworker Motivation Scale}
\label{app:scale}
\begin{table}[h!b]
\caption{\textbf{The Multidimensional Crowdworker Motivation Scale.} The stem is ``Why do you or would you put efforts into doing CrowdFlower tasks?'' (adapted from MWMS \cite{mwms}). All items were answered on a 7-point Likert scale ranging from ``not at all'' (1) to ``completely'' (7). The column ``Source'' indicates from which motivation scale the item was adapted.}
\centering
%\resizebox{\textwidth}{!}{
\begin{tabular}{p{15mm} l c} 
  & & \multicolumn{1}{c}{Source}\\
\toprule
  \textbf{Amotivation} & &\\
  \tab Am1&\it{I don't know why, CrowdFlower tasks often seem like a waste of time.}&\\ 
  \tab Am2&\it{I don't know why I'm doing CrowdFlower tasks, it's pointless work.} & \cite{mwms}\\
    \tab Am3&\it{I don't know why, I often perceive CrowdFlower tasks as an annoying chore.}& \cite{teaching}\\ [.5mm]
    
  \textbf{\mbox{External Regulation (Material)}} && \\
  \tab ExMat1&\it{Because CrowdFlower tasks give me financial gains.} &\\ 
  \tab ExMat2&\it{For the income CrowdFlower tasks provide me.} & \cite{weims}\\ 
  \tab ExMat3&\it{Because of the money I get from doing CrowdFlower tasks.} &\\[.5mm] 
  
  \textbf{\mbox{External Regulation (Social)}} && \\
  \tab ExSoc1&\it{Because other people want me to do CrowdFlower tasks (e.g. family, friends,...).} &\\ 
  \tab ExSoc2&\it{Because other people say I should (e.g. family, friends,...).} & \cite{walking}\\
  \tab ExSoc3&\it{Because other people expect it of me (e.g. family, friends,...).}& \\ [.5mm] 
  
  \textbf{\mbox{Introjected Regulation}} && \\
  \tab Introj1&\it{Because otherwise I would have a bad conscience.} & \cite{teaching}\\ 
  \tab Introj2&\it{Because otherwise I will feel ashamed of myself.} & \cite{mwms}\\ 
  \tab Introj3&\it{Because otherwise I will feel bad about myself.} & \cite{mwms}\\[.5mm] 
  
  \textbf{\mbox{Identified Regulation}} && \\
 \tab Ident1&\it{Because this is the type of work I chose to do to attain a certain lifestyle.} & \cite{weims}\\ 
 \tab Ident2&\it{Because I chose this type of work to attain my career goals.} & \cite{weims} \\ 
 \tab Ident3&\it{Because it is the type of work I have chosen to attain certain important objectives.} & \cite{weims} \\[.5mm] 
 
 \textbf{\mbox{Intrinsic Motivation}} && \\ 
 \tab Intrin1&\it{Because I have fun doing CrowdFlower tasks.} & \cite{mwms}\\  
 \tab Intrin2&\it{Because I enjoy doing CrowdFlower tasks.} & \cite{walking}\\ 
 \tab Intrin3&\it{Because what I do in CrowdFlower tasks is interesting.} & \cite{mwms} \\[.5mm] 
 \bottomrule
\end{tabular}
%}
\label{table:mcms_items}
\end{table}

\clearpage

\section{McDonald's Coefficient Omega}
\label{app:omega}

As an alternative to Cronbach's alpha, we additionally calculated McDonald's coefficient omega \cite{mcdonald1999test} for assessing the reliability of the MCMS. Table~\ref{table:mcdonalds_omega} displays the values of omega for each country and income group. 
As with Crohnbach's alpha, in most countries and groups, omega exceeds 0.7 for each construct. Exceptions to this are the amotivation factor in Brazil and Venezuela as well as the material external regulation factor in Brazil and Indonesia with values between 0.5 and 0.7. Compared to Crohnbach's alpha, the values of coefficient omega are equal or slightly higher.\footnote{One exception to this was intrinsic motivation in the high income group, where coefficient omega is 0.01 lower than alpha.} 

\begin{table}[b!]
\caption{\textbf{McDonald's Coefficient Omega.} This table shows McDonald's omega values for all groups and constructs, along with a 95\% confidence interval for the values.}
\centering
%\resizebox{\columnwidth}{!}{
\begin{tabular}{l c c c c c c}
\toprule
  \textbf{Group} & \textbf{Amotivation} & \textbf{Material} & \textbf{Social} & \textbf{Introjected} & \textbf{Identified} & \textbf{Intrinsic}\\
\midrule
ALL & \ci{ 0.77 }{ 0.78 }{ 0.8 } & \ci{ 0.78 }{ 0.79 }{ 0.81 } & \ci{ 0.83 }{ 0.84 }{ 0.85 } & \ci{ 0.82 }{ 0.83 }{ 0.84 } & \ci{ 0.86 }{ 0.87 }{ 0.88 } & \ci{ 0.88 }{ 0.88 }{ 0.89 } \\
\midrule
HIGH & \ci{ 0.82 }{ 0.84 }{ 0.86 } & \ci{ 0.8 }{ 0.82 }{ 0.84 } & \ci{ 0.84 }{ 0.86 }{ 0.88 } & \ci{ 0.84 }{ 0.86 }{ 0.88 } & \ci{ 0.85 }{ 0.87 }{ 0.88 } & \ci{ 0.88 }{ 0.89 }{ 0.91 } \\
MID & \ci{ 0.67 }{ 0.71 }{ 0.74 } & \ci{ 0.74 }{ 0.76 }{ 0.79 } & \ci{ 0.81 }{ 0.83 }{ 0.84 } & \ci{ 0.81 }{ 0.83 }{ 0.85 } & \ci{ 0.85 }{ 0.86 }{ 0.88 } & \ci{ 0.86 }{ 0.88 }{ 0.89 } \\
LOW & \ci{ 0.74 }{ 0.78 }{ 0.81 } & \ci{ 0.74 }{ 0.78 }{ 0.81 } & \ci{ 0.84 }{ 0.86 }{ 0.87 } & \ci{ 0.77 }{ 0.80 }{ 0.83 } & \ci{ 0.83 }{ 0.85 }{ 0.87 } & \ci{ 0.86 }{ 0.87 }{ 0.89 } \\
\midrule
USA & \ci{ 0.83 }{ 0.86 }{ 0.88 } & \ci{ 0.81 }{ 0.84 }{ 0.88 } & \ci{ 0.8 }{ 0.84 }{ 0.88 } & \ci{ 0.79 }{ 0.83 }{ 0.88 } & \ci{ 0.83 }{ 0.85 }{ 0.88 } & \ci{ 0.89 }{ 0.91 }{ 0.92 } \\
ESP & \ci{ 0.81 }{ 0.84 }{ 0.87 } & \ci{ 0.78 }{ 0.81 }{ 0.85 } & \ci{ 0.83 }{ 0.86 }{ 0.88 } & \ci{ 0.85 }{ 0.87 }{ 0.9 } & \ci{ 0.88 }{ 0.89 }{ 0.91 } & \ci{ 0.89 }{ 0.9 }{ 0.92 } \\
DEU & \ci{ 0.8 }{ 0.83 }{ 0.86 } & \ci{ 0.78 }{ 0.81 }{ 0.85 } & \ci{ 0.85 }{ 0.88 }{ 0.91 } & \ci{ 0.78 }{ 0.82 }{ 0.87 } & \ci{ 0.82 }{ 0.85 }{ 0.87 } & \ci{ 0.84 }{ 0.86 }{ 0.89 } \\
BRA & \ci{ 0.44 }{ 0.52 }{ 0.61 } & \ci{ 0.62 }{ 0.69 }{ 0.76 } & \ci{ 0.8 }{ 0.83 }{ 0.86 } & \ci{ 0.77 }{ 0.81 }{ 0.85 } & \ci{ 0.87 }{ 0.89 }{ 0.91 } & \ci{ 0.83 }{ 0.86 }{ 0.9 } \\
RUS & \ci{ 0.73 }{ 0.77 }{ 0.81 } & \ci{ 0.78 }{ 0.81 }{ 0.85 } & \ci{ 0.87 }{ 0.89 }{ 0.91 } & \ci{ 0.86 }{ 0.88 }{ 0.9 } & \ci{ 0.83 }{ 0.85 }{ 0.88 } & \ci{ 0.87 }{ 0.89 }{ 0.91 } \\
MEX & \ci{ 0.74 }{ 0.79 }{ 0.84 } & \ci{ 0.68 }{ 0.73 }{ 0.79 } & \ci{ 0.74 }{ 0.78 }{ 0.81 } & \ci{ 0.74 }{ 0.78 }{ 0.82 } & \ci{ 0.83 }{ 0.85 }{ 0.87 } & \ci{ 0.83 }{ 0.86 }{ 0.89 } \\
IND & \ci{ 0.68 }{ 0.74 }{ 0.8 } & \ci{ 0.78 }{ 0.82 }{ 0.86 } & \ci{ 0.84 }{ 0.87 }{ 0.89 } & \ci{ 0.75 }{ 0.79 }{ 0.83 } & \ci{ 0.83 }{ 0.85 }{ 0.88 } & \ci{ 0.85 }{ 0.88 }{ 0.9 } \\
IDN & \ci{ 0.74 }{ 0.79 }{ 0.85 } & \ci{ 0.57 }{ 0.65 }{ 0.72 } & \ci{ 0.83 }{ 0.86 }{ 0.89 } & \ci{ 0.74 }{ 0.79 }{ 0.84 } & \ci{ 0.74 }{ 0.79 }{ 0.84 } & \ci{ 0.81 }{ 0.85 }{ 0.89 } \\
PHL & \ci{ 0.75 }{ 0.81 }{ 0.86 } & \ci{ 0.75 }{ 0.8 }{ 0.85 } & \ci{ 0.81 }{ 0.84 }{ 0.87 } & \ci{ 0.77 }{ 0.82 }{ 0.86 } & \ci{ 0.86 }{ 0.88 }{ 0.91 } & \ci{ 0.86 }{ 0.89 }{ 0.91 } \\
VEN & \ci{ 0.52 }{ 0.63 }{ 0.74 } & \ci{ 0.72 }{ 0.77 }{ 0.83 } & \ci{ 0.74 }{ 0.77 }{ 0.81 } & \ci{ 0.73 }{ 0.77 }{ 0.82 } & \ci{ 0.84 }{ 0.86 }{ 0.89 } & \ci{ 0.8 }{ 0.84 }{ 0.87 } \\
\bottomrule
\end{tabular}
%}
\label{table:mcdonalds_omega}
\end{table}

\clearpage

\section{Task Interface}
\label{app:interface}

The CrowdFlower task included a section in which workers were instructed to write down five reasons for why they do tasks on CrowdFlower, a section with the MCMS, and a section with questions about demographics and money use.

\begin{figure}[h!]
	\centering
	\begin{center}
            \includegraphics[width=0.9\textwidth]{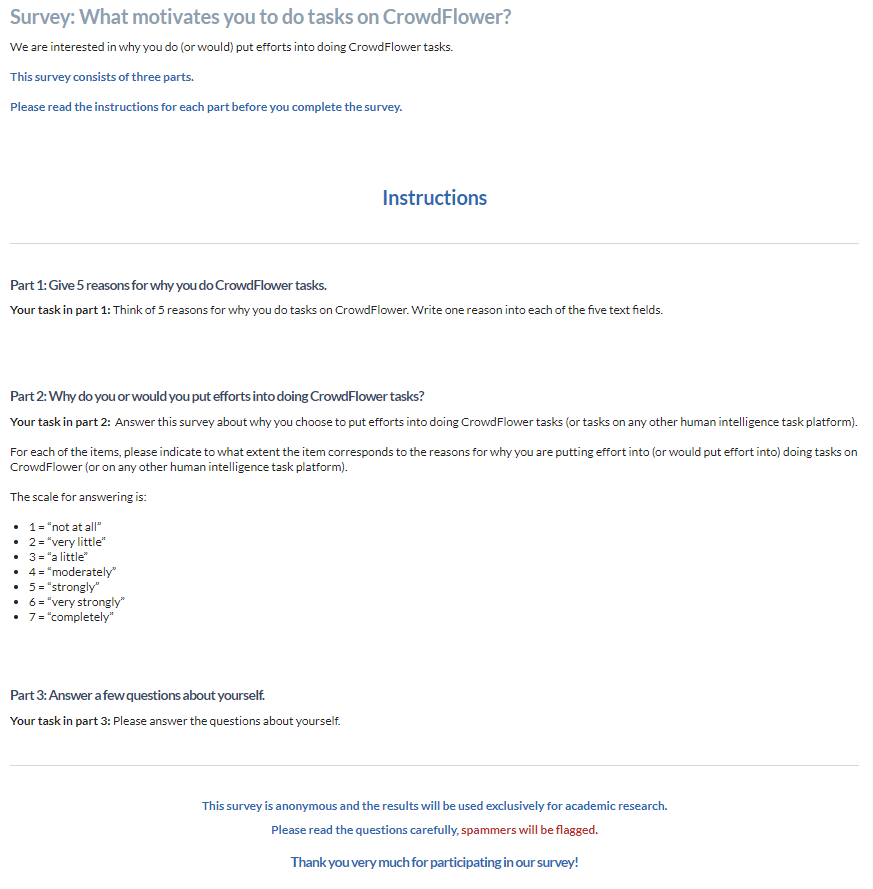}
	\end{center}	

  	\caption{\textbf{Task instructions.} This figure shows the task instructions that were shown to crowdworkers at the beginning of the task.}
  \centering
  \label{fig:task_interface_instructions_all}
\end{figure}

\begin{figure}[h!]
	\centering
	\begin{center}
            \includegraphics[width=0.8\textwidth]{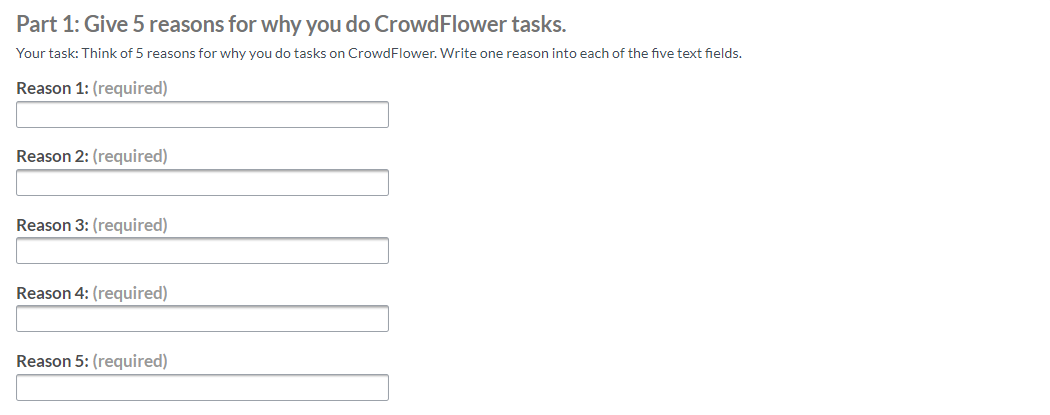}
	\end{center}	

  	\caption{\textbf{Interface for open-ended answers.} This figure shows the interface that crowdworkers were given to state five reasons for why they do tasks on CrowdFlower in their own words.}
  \centering
  \label{fig:task_interface_reasons}
\end{figure}

\begin{figure}[h!]
	\centering
	\begin{center}
            \includegraphics[width=0.8\textwidth]{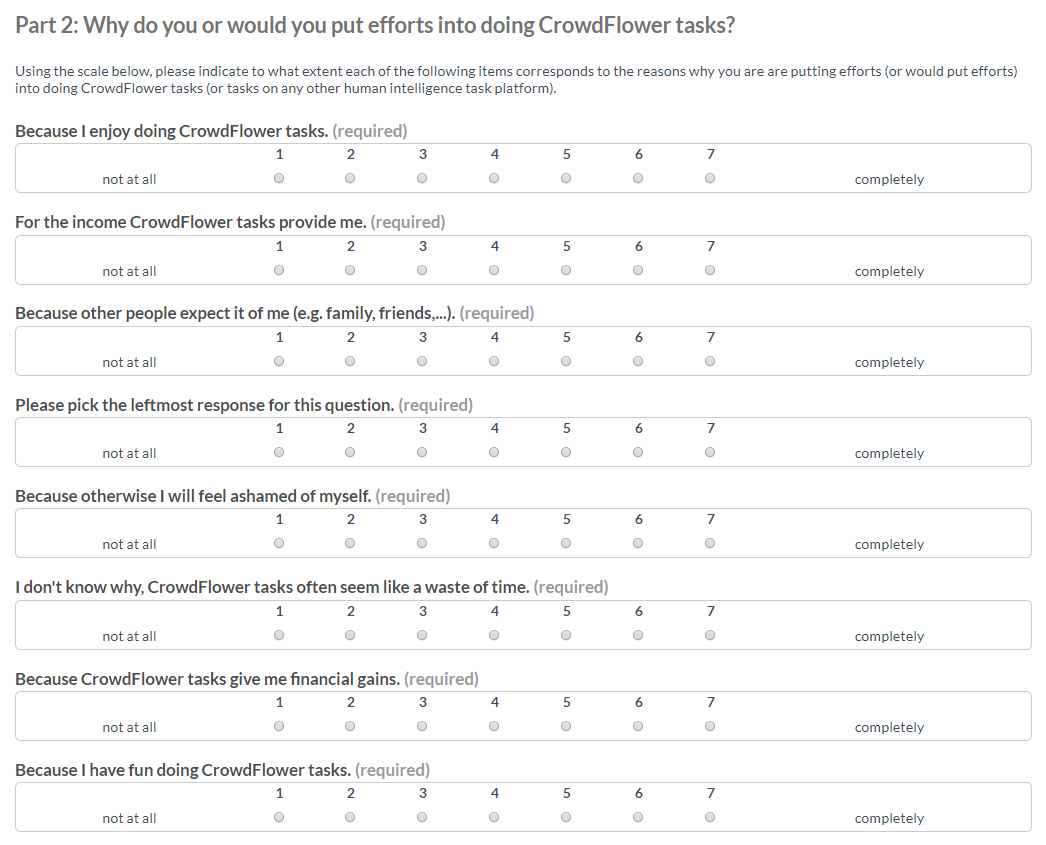}
	\end{center}	

  	\caption{\textbf{MCMS Interface.} This figure shows the interface of the task section in which crowdworkers answered the MCMS question by indicating agreement on the 18 items. Also note the included test question to check for spamming behavior and lack of attention. Due to space limitations, this screenshot shows only part of the scale. The full scale is shown in Appendix~\ref{app:scale}.}
  \centering
  \label{fig:task_interface_mcms}
\end{figure}

\clearpage

\section{Instructions for Use of the MCMS}
\label{app:scale_use}

To measure the motivation of crowdworkers, researchers can administer the scale (shown in Appendix~\ref{app:scale}) to their sample of workers. For example, the MCMS can be easily included as a module in the design of a micro task if a researcher wishes to measure motivation along with other variables, such as other characteristics or behavioral patterns, of their worker sample.

As long as the researcher does not wish to compare different groups of crowdworkers, the MCMS can be used as a summated scale. This means that the scores for the constructs can be obtained by averaging the scores of the individual items corresponding to each construct. For example, to obtain the score for the amotivation construct of a specific worker, the researcher takes the individual item responses of this worker for the items associated with the amotivation construct (i.e., Am1, Am2 and Am3) and calculates $Amotivation = (Am1 + Am2 + Am3)/3.0$.

An alternative way to obtain the construct scores is to specify the model shown in Figure~\ref{fig:sem_diagram} as a latent variable model, for example by using latent variable modeling software such as Mplus\footnote{https://www.statmodel.com/} or the R library lavaan\footnote{https://cran.r-project.org/web/packages/lavaan/lavaan.pdf}. This method\footnote{For an introduction to latent variable models and multivariate data analysis, we refer the reader to Kline \cite{kline2015principles}, Hair et al. \cite{hair2018multivariate} and Bollen \cite{bollen1989sem}.} of obtaining the construct scores is preferable to averaging manifest item scores as it accounts for the measurement error necessarily present in the measurement of any abstract concept \cite{hair2018multivariate}. Furthermore, specifying the model allows researchers to conduct confirmatory factor analysis to validate the MCMS on their sample of crowdworkers, ensuring that the measurement is valid for their specific target population.

If a researcher wishes to compare the construct means between two groups of crowdworkers, such as male and female workers, workers of different age, workers on different platforms, or, as in the case of the present study, different countries, the researcher has to ensure that the workers of the different groups assign the same meaning to the items used in the MCMS. This can be done by establishing measurement invariance between the groups as described in Section~\ref{sec:invariance}. Measurement invariance should be established before comparing the group means of any scale, especially when administering the scale to respondents of different cultures, to ensure that the differences in scores are not due to a different understanding of the items (e.g. due to culture) or due to measurement artifacts such as different levels of extreme or acquiescent response bias (e.g. \cite{cheung2000assessing}).

\end{document}